\documentclass[10pt,letterpaper]{article}
\usepackage[top=0.85in,left=1.5in,footskip=0.35in]{geometry}

\usepackage{amsmath,amssymb}

\usepackage{changepage}

\usepackage{textcomp,marvosym}

\usepackage{cite}

\usepackage{nameref,hyperref}

\usepackage[right]{lineno}

\usepackage[nopatch=eqnum]{microtype}
\DisableLigatures[f]{encoding = *, family = * }

\usepackage[table]{xcolor}

\usepackage{array}

\newcolumntype{+}{!{\vrule width 2pt}}

\newlength\savedwidth

\raggedright
\usepackage[aboveskip=1pt,labelfont=bf,labelsep=period,justification=raggedright,singlelinecheck=off]{caption}

\makeatletter
\renewcommand{\@biblabel}[1]{\quad#1.}
\makeatother

\usepackage{lastpage,fancyhdr,graphicx}
\usepackage{epstopdf}
\usepackage{ragged2e}
\fancyheadoffset[L]{1.25in}
\fancyfootoffset[L]{1.25in}
\begin{document}
\vspace*{0.2in}
\justifying

\begin{flushleft}
{\Large
\textbf\newline{M-SDT: A modelling framework for dengue transmission, forecasting, and intervention strategies in Ahmedabad Municipal Corporation} 
}
\newline
\\
Sourav Roy\textsuperscript{1},
Rajendra Gadhavi\textsuperscript{1},
Bhavin Solanki\textsuperscript{2},
Chirag Shah\textsuperscript{2},
Raj C. Sharma\textsuperscript{2},
Indrajit Ghosh\textsuperscript{1*},

\bigskip
\textbf{1} Bagchi School of Public Health, Ahmedabad University, Gujarat, India
\\
\textbf{2} Ahmedabad Municipal Corporation, Ahmedabad, Gujarat, India
\bigskip

%
%





* Corresponding author (indrajit.ghosh@ahduni.edu.in)

\end{flushleft}
\section*{Abstract}
Dengue fever poses a persistent public health challenge in rapidly urbanizing Indian cities such as Ahmedabad, where spatial heterogeneity and seasonal variability complicate forecasting and control. In this study, we develop a data-driven compartmental framework to simulate transmission dynamics, generate forecasts, and evaluate intervention strategies across the Ahmedabad Municipal Corporation (AMC). We employ a Mechanistic Seasonal Dengue Transmission (M-SDT) model that incorporates symptomatic and asymptomatic infections. We calibrated the proposed model using zone-wise dengue case data from aggregate public and private healthcare systems during 2020--2024. Parameter uncertainty is rigorously quantified using a bootstrap sampling framework with negative binomial noise that can explicitly account for observational overdispersion. The calibrated model reveals pronounced spatial heterogeneity across AMC zones, with persistent hotspots and distinct transmission regimes. Forecasts for 2026--2028 indicate continued endemic circulation with moderate inter-annual variability. Sensitivity analysis identifies the mosquito biting rate and vector mortality as dominant drivers of long-term disease burden, highlighting the central role of vector ecology in shaping epidemic outcomes. Evaluating seasonal vector control strategies shows a notable difference in operation; periodic fogging has a cumulative effect over the years, while sustained residual spraying can quickly curb outbreaks and decrease incidence by over 80\%. The overall effectiveness of these control mechanisms are influenced by the intensity of local transmission. The zone-wise analysis reveals that the mosquito-to-human ratio governs not only the baseline outbreak potential but also each zone's responsiveness to control strategies. Overall, the M-SDT modelling framework enables reconstruction of unobserved dynamics, rigorous uncertainty quantification, and evaluation of targeted, zone-specific interventions, underscoring the importance of integrating fine-scale surveillance data with mechanistic modelling for adaptive urban dengue control.


\nolinenumbers

\section{Introduction}
Dengue fever is a mosquito-borne disease that is exerting a substantial and growing burden on public health systems in tropical and subtropical regions \cite{zhang2025assessing}. Dengue infection, mainly transmitted by Aedes aegypti, presents a broad spectrum of clinical manifestations, ranging from mild fever to severe clinical symptoms, such as dengue haemorrhagic fever and dengue shock syndrome \cite{jing2019dengue}. Dengue infection, mainly transmitted by Aedes aegypti, shows a wide spectrum of clinical presentations, from mild fever to severe forms such as dengue haemorrhagic fever and dengue shock syndrome \cite{jing2019dengue}. Recent estimates show that hundreds of millions of infections occur annually, and almost half of the world’s population is at risk \cite{bhatt2013global, stanaway2016global}. The incidence of dengue has increased dramatically around the world in recent decades, with the number of cases reported to the World Health Organization (WHO) increasing from 5,05,430 cases in 2000 to 14.6 million in 2024. In 2024, India recorded the highest number of dengue cases with 2,33,519 cases and 297 deaths reported till the end of the year, while Gujarat reported 7,891 cases and 6 deaths \url{https://ncvbdc.mohfw.gov.in/}. Repeated seasonal outbreaks have been observed in Ahmedabad with a significant rise in reported cases during post-monsoon periods of 2022--2025. These patterns demonstrate the increasing public health burden at national and regional levels.

The burden of dengue is especially evident in areas that are rapidly being urbanized, where high population density, environmental heterogeneity, and inadequate vector control create ideal conditions for continued transmission. India is one of the major dengue transmission hotspots and several major cities have seen frequent outbreaks of dengue. Cities such as Ahmedabad experience large inter-annual variability in dengue incidence, driven by a complex interplay of climatic forcing, demographic growth and heterogeneous vector ecology \cite{gupta2012dengue, shepard2016economic}. In particular, the strong seasonality of the monsoon cycle critically affects transmission intensity, influencing both mosquito abundance and human-vector contact rates. These features highlight the importance of developing region-specific mechanistic models that are able to incorporate both environmental forcing and local epidemiological structure. One of the main challenges in dengue epidemiology is the high proportion of asymptomatic and mild symptomatic infections that are not usually captured by routine surveillance systems. These empirical investigations have shown that such infections represent a significant portion of all infections and may play an important role in causing secondary infections \cite{duong2015asymptomatic, yoon2012underrecognized}. This hidden infection source makes it difficult to determine the burden of illness, which makes control efforts based only on documented cases ineffective. As a result, models that account for asymptomatic transmission play a crucial role in accurately modelling the spread of epidemics \cite{ghosh2021reservoir}.

Models of dengue disease transmission that use mathematical techniques have given significant information about epidemic behavior and control, generally relying on coupled human–mosquito compartmental systems \cite{esteva1998model, ferguson1999dynamics}. Recent developments in model building for the purpose of better prediction have focused on making models more realistic by including aspects like seasonality, climate dependency, and data-informed parameter estimation \cite{lourenco2014dengue, siraj2017temperature, ghosh2019effect, aguiar2022mathematical}. At the same time, there has been increased emphasis on the inclusion of spatial heterogeneity, human movement, and multiple strains for improved understanding of disease dynamics~\cite{stoddard2009role, wesolowski2015impact, reiner2016quantifying, senapati2019impact}. Other studies have highlighted the role played by asymptomatic and subclinical infections in maintaining hidden sources of transmission, potentially affecting the scale and timing of outbreaks \cite{duong2015asymptomatic}. In addition, new approaches to data assimilation and inferential modelling have facilitated enhanced characterisation of transmission dynamics from surveillance data, but may still suffer from a lack of spatial detail or mechanistic clarity \cite{shaman2013real, funk2019assessing}.

Nevertheless, most previous studies either assume homogeneous disease transmission with constant parameters, exclude asymptomatic infections, or lack epidemiological data for the particular area in question \cite{kumar2024mathematical, smith2012ross}. Furthermore, some epidemiological models do not account for possible under-reporting of observed cases \cite{ghosh2019effect}. Therefore, their applicability to managerial decision-making on a practical level is limited. Further, the efficacy of widely used vector control measures, such as fogging and residual spraying, depends strongly on the timing, frequency, and seasonality of these interventions. Although widely used, there is currently no quantitative, mechanistic evaluation of the interaction between the above control methods and the underlying transmission dynamics, especially when these processes are strongly driven by seasonality and involve unknown reservoirs. Such research will require models capable of explicitly modelling both the epidemiology and the intervention process over time.

The proposed model is an SDT-SEIAR-mosquito-based modelling approach, referred to as M-SDT, to analyze dengue virus transmission in AMC and the seven constituent areas. Specifically, the proposed model accounts for asymptomatic infection cases, mosquito seasonality, and coupling between the human and vector populations. We use dengue incidence data from 2020 to 2024 to calibrate the model and calculate the transmission parameters using bootstrapped inference methods \cite{chowell2017fitting}. Short-term forecasts of the number of dengue cases for the years 2026 to 2028 are developed using the fitted model. Further expanding the model to consider the above-mentioned results, we include time-dependent interventions that reflect the effects of fogging and spraying to alter mosquito mortality rates within the modelling framework.

By combining mechanistic models, statistical inference, and intervention analysis, this paper presents a comprehensive framework for understanding dengue transmission in an urban context. The M-SDT framework will also provide useful quantitative insights into the design of optimal vector control measures in realistic epidemiological settings.

The major contributions of the current study can be highlighted as follows:
\begin{enumerate}
    \item A transmission dynamics model, M-SDT, has been established by fitting it to dengue incidence data from AMC and its seven regions.
    \item The M-SDT model provides a timely forecast of dengue incidence in AMC and its seven regions.
    \item Time-dependent intervention strategies representing fogging and spraying are examined under plausible scenarios in AMC and its seven regions.
\end{enumerate}

\begin{figure} [h]
    \centering
    \includegraphics[width=1.0\linewidth]{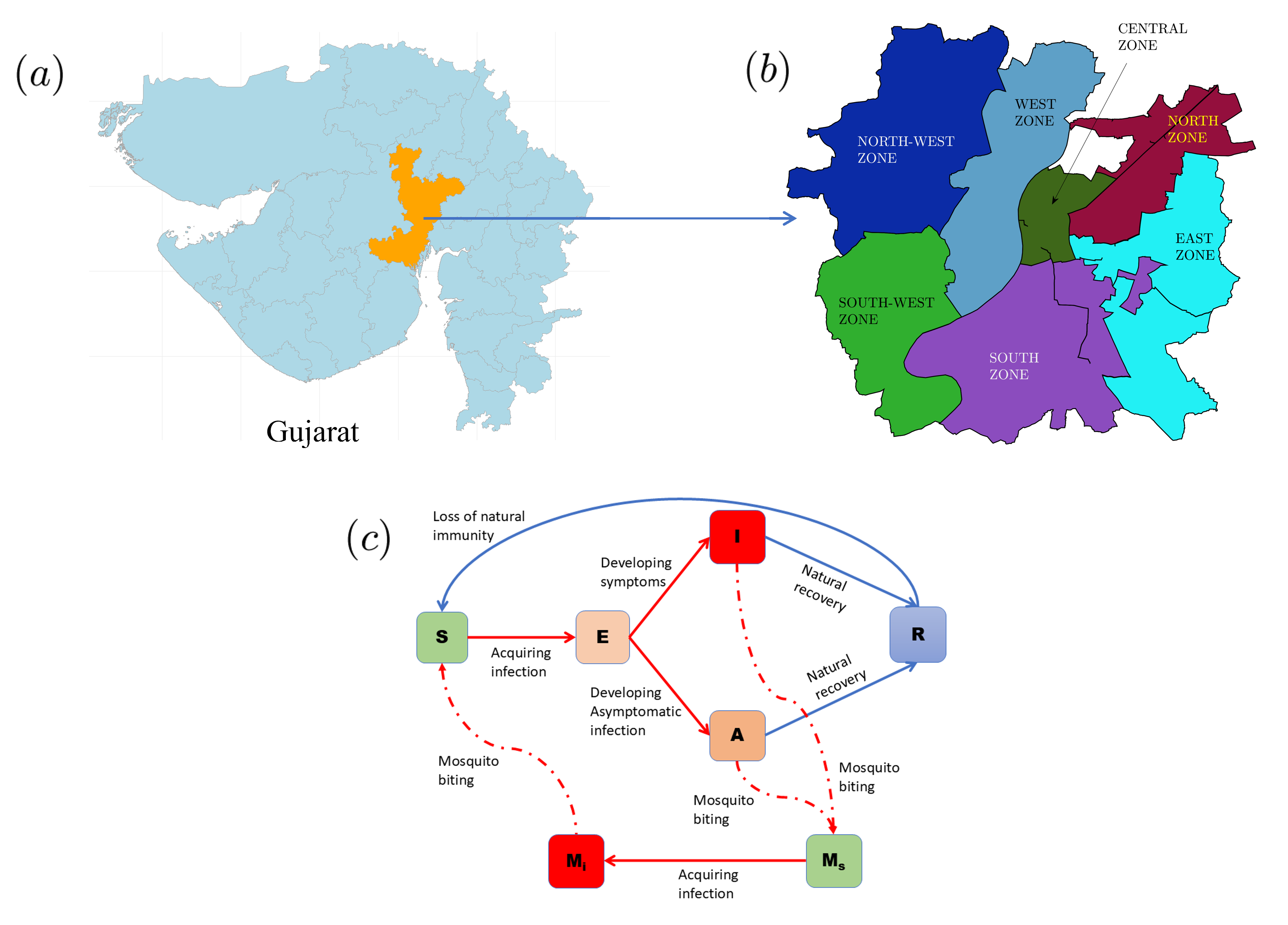}
    \caption {($a$) Map of Gujarat highlighting the Ahmedabad District. ($b$) Map of AMC divided into seven regions of study. ($c$) Schematic representation of the M-SDT model. The human population is divided into susceptible ($S$), exposed ($E$), symptomatic infected ($I$), asymptomatic infected ($A$), and recovered ($R$) compartments. Infection is acquired through bites from infected mosquitoes ($M_I$), while susceptible mosquitoes ($M_S$) become infected after biting infectious humans ($I$ and $A$). Exposed individuals progress to either symptomatic or asymptomatic states, with asymptomatic individuals having reduced infectiousness. Recovered individuals lose immunity over time and return to the susceptible class. Solid arrows represent epidemiological transitions, while dashed arrows denote vector-mediated transmission pathways.}
    \label{fig:flow_diagram}
\end{figure}

The rest of this paper is organized as follows. The formulation of the M-SDT model along with its epidemiological assumptions is discussed in Section \ref{Sec_Model_formulation}. The results are presented in Section \ref{sec_results}, which includes the sensitivity analysis, parameter estimation, forecasting, and interventions. Finally, the discussion and conclusion section can be found in Section \ref{sec_discussion_conclusion}.

\section{Model Description}\label{Sec_Model_formulation}
To describe the transmission dynamics of dengue in the AMC and regions therein (Fig.~\ref{fig:flow_diagram}(b)), we formulate an M-SDT model that captures the coupled interactions between human hosts and \textit{Aedes} mosquito vectors. A schematic representation of the model structure is shown in Fig.~\ref{fig:flow_diagram}(c).

The human population at time $t$ is divided into five epidemiological compartments: susceptible $S(t)$, exposed $E(t)$, symptomatic infected $I(t)$, asymptomatic infected $A(t)$, and recovered $R(t)$. Susceptible individuals acquire infection through bites from infected mosquitoes and move into the exposed class. Following an intrinsic incubation period, exposed individuals progress to either a symptomatic or an asymptomatic infectious state. Both symptomatic and asymptomatic individuals contribute to transmission, with asymptomatic individuals exhibiting reduced infectiousness. Recovery is made by individuals in both classes and leads to the transition into the recovered class. Individuals from the recovered class rejoin the susceptible class as immunity wanes. Mosquitoes are categorized into two compartments: susceptible mosquitoes denoted by $M_S(t)$ and infected mosquitoes represented by $M_I(t)$. Susceptible mosquitoes are infected when they bite infectious humans, both symptomatic and asymptomatic. The infection of susceptible humans occurs once the infected mosquitoes bite them. The two-way process involving humans and vectors makes up the basic feedback loop for dengue transmission.

One notable characteristic of dengue transmission in AMC is its seasonal behaviour, driven by environmental changes during the monsoon, such as rainfall, temperature, and humidity, which affect mosquito breeding and biting rates \cite{enduri2017estimation}. To capture this, the biting rate per mosquito is described as a time-varying periodic function~\cite{hartley2002seasonal}. To incorporate this mathematically, we model the per-mosquito biting rate as a time-dependent periodic function~\cite{hartley2002seasonal}
\begin{equation}
b(t) = b_{0}\left(1 + a_{b}\sin\left(\frac{2\pi t}{365}\right)\right),
\label{eq:bt}
\end{equation}
where $b_{0}$ denotes the baseline biting rate and $a_{b} \in [0,1)$ represents the amplitude of seasonal forcing.

This time-dependent biting rate in Eq.~\eqref{eq:bt} determines the forces of infection for both humans and mosquitoes, which are given by
\begin{equation}
\lambda_H(t) = \frac{b(t)\,\beta_{HM}\,M_I}{N_H}, 
\qquad
\lambda_M(t) = \frac{b(t)\,\beta_{MH}\,\big(I + \theta A\big)}{N_H},
\label{eq:foi_seasonal}
\end{equation}
where $N_H = S + E + I + A + R$ represents the total population of humans, while $\theta \in [0,1]$ measures the decreased infectivity of asymptomatic humans. To maintain demographic balance, the human recruitment rate can be stated as proportional to the population size, that is, $\Pi_H = b_H N_H$, where $b_H$ is the human birth rate per capita.

Thus, the full dynamical system describing dengue dynamics is expressed as
\begin{equation} \label{eq:model}
\begin{array}{lcl}
\displaystyle \frac{dS}{dt} = \Pi_H - (\lambda_H + \mu_H)S + \omega_R R, \\[10pt]
\displaystyle \frac{dE}{dt} = \lambda_H S - (\sigma + \mu_H)E, \\[10pt]
\displaystyle \frac{dI}{dt} = p\sigma E - (\gamma_I + \mu_H + \xi)I, \\[10pt]
\displaystyle \frac{dA}{dt} = (1-p)\sigma E - (\gamma_A + \mu_H)A, \\[10pt]
\displaystyle \frac{dR}{dt} = \gamma_I I + \gamma_A A - (\omega_R + \mu_H)R, \\[10pt]
\displaystyle \frac{dM_S}{dt} = \mu_M N_M - \lambda_M M_S - \mu_M M_S, \\[10pt]
\displaystyle \frac{dM_I}{dt} = \lambda_M M_S - \mu_M M_I,
\end{array}
\end{equation}
with $N_M = M_S + M_I$ representing the mosquito population.

A complete listing of all the model parameters, along with their biological significance, units, and standard ranges is provided in Table~\ref{table1}. The majority of these parameter values have been taken from literature on dengue models for India and similar contexts.
\begin{table}[h!]  
\centering
\footnotesize
\begin{tabular}{lp{3cm}lll}
\hline
Parameter & Parameter meaning & Units & Typical value / range & Reference \\
\hline
$N_H$ & Total human population & individuals & region-specific & \cite{senapati2019impact,ghosh2019effect} \\
$N_M$ & Total mosquito population & individuals & $mN_H$, $m \in [1,5]$ & \cite{senapati2019impact} \\[3pt]

$\mu_H$ & Human natural mortality rate & day$^{-1}$ & $3\times 10^{-5}$ & \cite{Gujarat1972} \\
$b_H$ & Per capita birth rate & day$^{-1}$ & $7.92 \times 10^{-5}$ & \cite{Gujarat1972} \\
$\sigma$ & Transition rate from Exposed & day$^{-1}$ & $3.3/365$ & \cite{pinho2010modelling} \\
$p$ & Fraction of infections that become symptomatic & dimensionless & [0,1] & Free parameter \\
$\gamma_I$ & Recovery rate (symptomatic) & day$^{-1}$ & $0.2$ & \cite{hamdan2021development} \\
$\gamma_A$ & Recovery rate (asymptomatic) & day$^{-1}$ & $1/4$ & \cite{ghosh2019effect, enduri2017estimation} \\
$\xi$ & Disease-induced mortality rate & day$^{-1}$ & very small ($10^{-5} - 10^{-4}$) & \cite{shepard2014economic} \\
$\omega_R$ & Waning immunity rate & day$^{-1}$ & $0.00137$ & \cite{katzelnick2017immune} \\[3pt]
$\mu_M$ & Mosquito mortality rate & day$^{-1}$ & $0.02 - 0.025$ & \cite{sardar2015mathematical} \\
$b_0$ & Biting rate per mosquito & day$^{-1}$ & [0,1] & Free parameter \\
$a_b$ & Amplitude with seasonal forcing & dimensionless & 0.4 - 0.6 & \cite{zahid2023biting}\\
$\beta_{HM}$ & Mosquito-to-human transmission probability & dimensionless & $0.75$ & \cite{newton1992model} \\
$\beta_{MH}$ & Human-to-mosquito transmission probability & dimensionless & $0.75$ & \cite{newton1992model} \\
$\theta$ & Relative infectivity of asymptomatic & dimensionless & $0.8$ & \cite{ten2018contributions} \\
\hline
\end{tabular} 
\caption{Model parameters for the SEAIR--mosquito system defined in Eq.~\eqref{eq:model}.}
\label{table1}
\end{table}

\subsection{Free and estimated parameters}

Despite most of the parameters being drawn from the literature, the remaining parameters are considered free parameters owing to their high local uncertainty and are calibrated based on data.

The initial baseline biting rate $b_0$ represents the impact of mosquito population and climatic factors together and its variation range is considered as $b_0 \in [0.01,\,1]\ \text{day}^{-1}$. The symptomatic fraction $p$ indicates the proportion of the clinical detection of an infection in the population and has a variation range of $p \in [0,\,1]$. The quantity $m$ denoting the relation between $N_M$ and $N_H$, i.e. $N_M = m N_H$, is considered for values within $m \in [1,\,5]$. It represents environmental conditions and the number of vectors and is based on a reporting rate of 5\% due to an expected significant degree of underreporting of cases, previously mentioned in the literature \cite{undurraga2013expansion, das2017underreporting}.

The combination of these parameters defines the transmission intensity and reporting system of dengue in AMC, and the calculation of their values makes the model able to replicate epidemiological observations and still be biologically interpretable. Finally, to maintain consistency with long-term demographic and epidemiological trends, the model is initialized with population data corresponding to 1960 and numerically integrated using the Runge-Kutta-45 method. This baseline allows the system to evolve before the observational period used for calibration and forecasting, thereby providing a coherent foundation for the results presented in the following sections.

\section{Results}\label{sec_results}
Global sensitivity analysis of the crucial model parameters is studied with respect to an epidemiologically relevant response variable. Building on the observations, the M-SDT model~\eqref{eq:model} is calibrated to the reported data using a nonlinear least-squares approach, enabling reconstruction of the underlying epidemiological processes. Subsequently, to incorporate variability and uncertainty in the reported cases, a bootstrap-based framework is employed to generate an ensemble of parameter estimates and corresponding model trajectories. We also examined the identifiability of the estimated parameters at this step. The calibrated model is then used to forecast future dengue incidence and to assess the effectiveness of different intervention strategies. This sequential approach, associating data analysis, model fitting, uncertainty quantification, and predictive assessment, gives a comprehensive understanding of dengue dynamics and supports informed public health decision-making.

\subsection{Sensitivity analysis}
In order to study the impact that different values for each of the parameters of our model had on the transmission dynamics of dengue, a global sensitivity analysis was carried out using the Partial Rank Correlation Coefficient method as described by Marino et al. \cite{marino2008methodology}.

\begin{figure}
    \centering
    \includegraphics[width=1.0\linewidth]{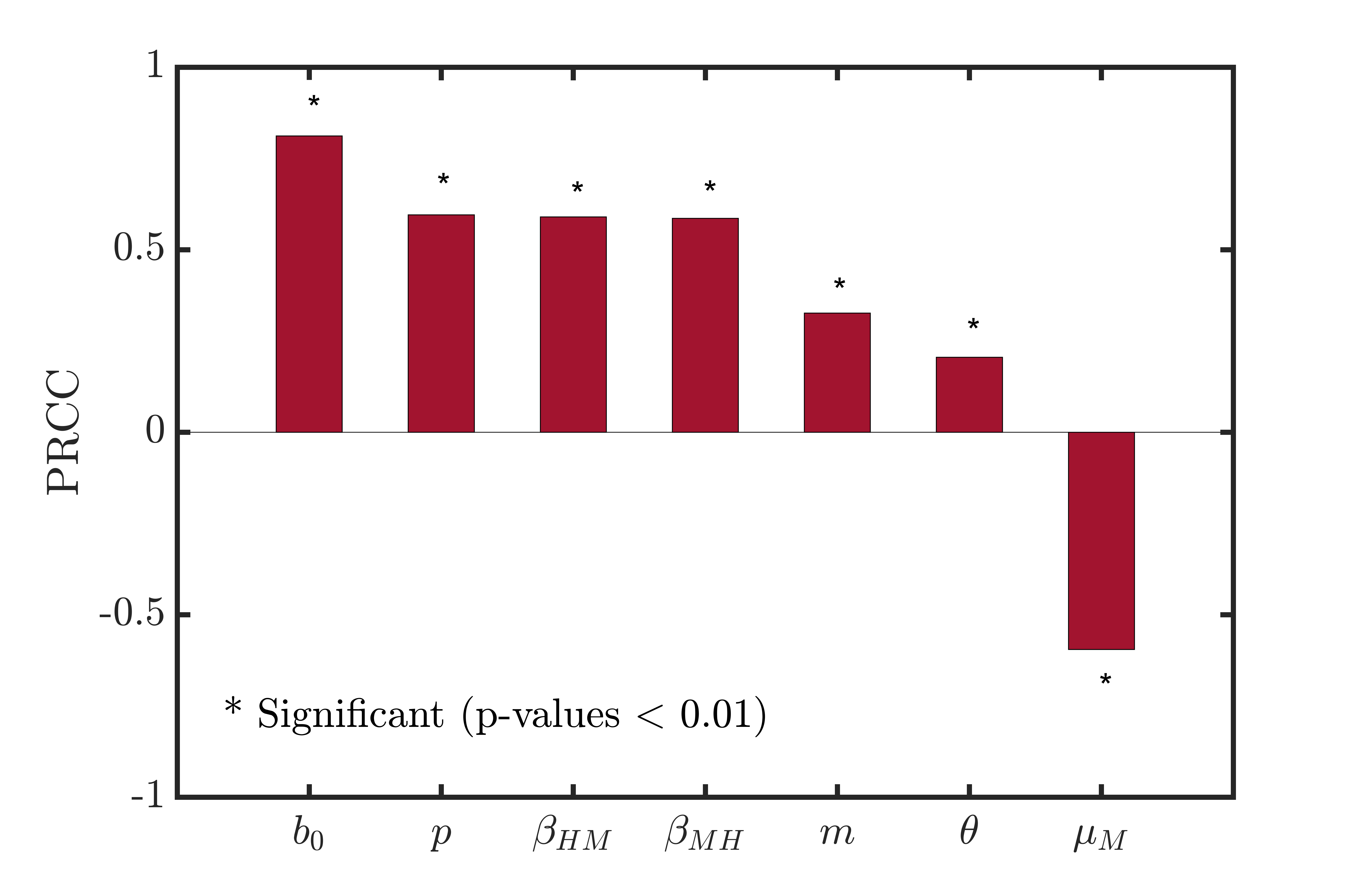}
    \caption{Global sensitivities of the model parameters to the cumulative symptomatic infections for the period 2026-2030 at AMC. Notable PRCC values are marked by \textbf{*} (p-value $<$ 0.01). Parameters that remain fixed are considered from Table~\ref{table1}.}
    \label{fig:PRCC_dengue}
\end{figure}

The variables analyzed included seven important input parameters: the biting rate of mosquitoes ($b_0$), probability of transmission from mosquitoes to humans ($\beta_{HM}$), probability of transmission from humans to mosquitoes ($\beta_{MH}$), the fraction of symptomatic infections ($p$), the relative infectivity of asymptomatic cases ($\theta$), vector to host ratio ($m$), and mosquito mortality rate ($\mu_M$). The cumulative projected number of dengue cases between the years 2026 and 2030 ($I_{\text{total}}^{\text{2026-2030}}$) was chosen as the output variable in this study. All parameters had a baseline range between (0,1) except for $m$ and $\theta$, which fell between (1,5) and (0,0.8), respectively. All other fixed parameters are selected from \ref{table1}. For systematically estimating the impact of individual parameters, the Latin Hypercube Sampling (LHS) sampling technique was adopted to generate 1,000 different sets of parameter values based on their respective probability distribution, enabling effective coverage of the multi-dimensional parameter space. The PRCC measures were calculated using the MATLAB package developed by Marino et al. \cite{marino2008methodology}, and the results are shown in Fig.~\ref{fig:PRCC_dengue}. The statistical significance level is set as $p < 0.01$.

These results show that there is a hierarchy among parameter effects. Among these, the rate at which mosquitoes bite people $b_0$ displays the highest positive association with the total incidence after 5 years, $I^{\mathrm{2026-2030}}$, as indicated in Fig.~\ref{fig:PRCC_dengue}. As this plays a dominant role, it demonstrates the critical role of human-vector contact in maintaining transmission. This is followed in descending order of effect by the fraction of symptomatic individuals $p$, the probability of transmission $\beta_{HM}$, $\beta_{MH}$, the ratio of vectors to hosts $m$, and the infectiousness of asymptomatic individuals $\theta$~\cite{correia2001new}.

In this respect, the sensitivity of $p$, which is relatively higher, shows that the disease burden in terms of the number of infected people can be significantly affected by the percentage of clinical cases. Also, the similar contribution of both $\beta_{HM}$ and $\beta_{MH}$ can point to the mutual interactions between humans and vectors in causing disease outbreaks. Hence, both the infectiousness of hosts and the competence of the vector should be considered. Lastly, $m$ describes the transmission amplification by the vector population, and thus, the significance of demographic and environmental factors can be recognized from its positive correlation. The influence of $\theta$, even though less than $\theta^*$, reveals that asymptomatic persons also contribute to the chain of transmission in a relatively non-trivial manner. On the other hand, the mosquito mortality rate ($\mu_M$) shows a substantial negative correlation with the response variable. The relationship is illustrated in Fig.~\ref{fig:PRCC_dengue} and shows the importance of vector longevity to epidemic persistence.

From the perspective of epidemiology, these results imply that any increase in the positively correlated parameters ($b_0$, $p$, $\beta_{HM}$, $\beta_{MH}$, $m$, and $\theta$) will result in a significant rise in the number of dengue cases in the coming five years. On the other hand, measures capable of increasing the vector death rate ($\mu_M$) will prove to be a key factor in reducing future dengue cases. Notably, the high sensitivity of $b_0$ and $\mu_M$ implies that vector management practices aimed at reducing mosquito population size and longevity, such as fumigation and residual spraying, can be highly effective in controlling disease prevalence.

\subsection{Model fitting and bootstrap-based parameter estimation}
The model~\eqref{eq:model} (M-SDT) is then estimated using the observed data on dengue incidence, using the methodology outlined in Section~\ref{methods}. The ratio between mosquitoes and humans ($m$) is set at $3.098$, based on long-term equilibrium estimation, thereby allowing for identifiable estimation of transmission parameters during recent outbreaks.

\begin{figure}[h]
\centering
\includegraphics[width = 1.0 \linewidth]{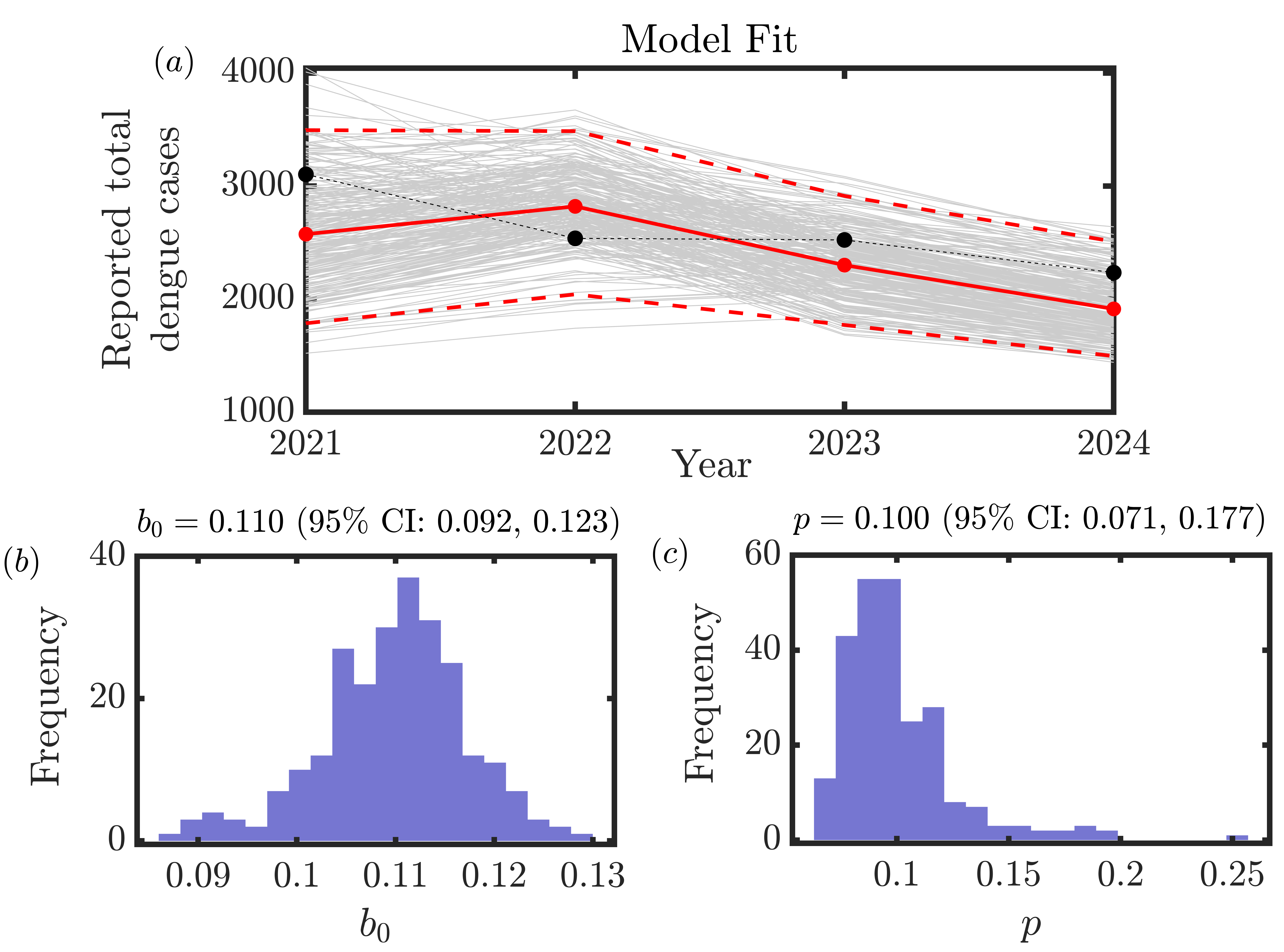} 
\caption{\textbf{Bootstrap uncertainty quantification and parameter inference in the M-SDT model:} Bootstrap-based uncertainty analysis of dengue transmission dynamics in AMC.
($a$) Empirical distribution of the biting rate $b_0$ obtained from 250 bootstrap realizations, with the median and 95\% confidence interval indicated.
($b$) Empirical distribution of the symptomatic fraction $p$ with corresponding 95\% confidence interval.
($c$) Annual reported dengue cases for 2021-2024. Black dots denote observed data, the solid red curve represents the median model prediction, and dashed red curves indicate the 95\% uncertainty envelope derived from bootstrap simulations. Gray curves show individual bootstrap realizations. The mosquito-to-human ratio was fixed based on calibration to the 2020 data, while $b_0$ and $p$ were estimated using annual incidence data from 2021 to 2024.}
\label{bootstrap}
\end{figure}

The model fit obtained is depicted in Fig.~\ref{bootstrap}($a$). Black circles denote the number of reported dengue cases per year from 2021 to 2024, whereas the red line illustrates the median of the model predictions for all bootstrap replications. Red dashed lines mark the $95\%$ confidence interval, and gray lines illustrate the individual fits of the bootstrap process. The high resemblance between the observed and the predicted values implies an adequate representation of the magnitudes and decreasing trends of the reported dengue cases. Crucially, the small variation in gray lines confirms accurate modelling of the transmission dynamics. With respect to the epidemic curve of the disease, the high value recorded in 2021 and the gradual reduction in later years fit well with post-epidemic trends seen in areas endemic for dengue infections. This is likely due to the temporary reduction in the number of susceptible individuals, the build-up of immunity in the population, and vector control programs carried out in response to major epidemics.

Distributions of the estimated parameters are shown in Figs.~\ref{bootstrap}($b$) and ($c$). The baseline biting rate is narrowly distributed around $b_0=0.110$ and its $95\%$ confidence interval is $[0.092,\,0.123]$. Such an interval for the estimated parameter can be interpreted as strong constraints on the transmission efficiency by the data and underlines the predominant importance of the vector--host interaction in determining the outbreak magnitude. In contrast, the confidence interval for the symptomatic fraction $p=0.100$ is wider $[0.071,\,0.177]$; see Fig.~\ref{bootstrap}($c$). Nevertheless, the histogram shows that the parameter is identifiable.

In summary, the confidence intervals from the bootstrap approach do not only offer an estimation of the statistical uncertainty but also reveal information regarding the epidemiology of dengue disease. As a result, the stability of $b_0$ implies that there are similar transmission rates across years, while the variation of $p$ indicates different levels of accessibility to health facilities and testing. In general, the collection of all possible parameter values generated from this method serves as the foundation for the forecast process.

\begin{figure}[h]
\centering
\includegraphics[width = 1.0 \linewidth]{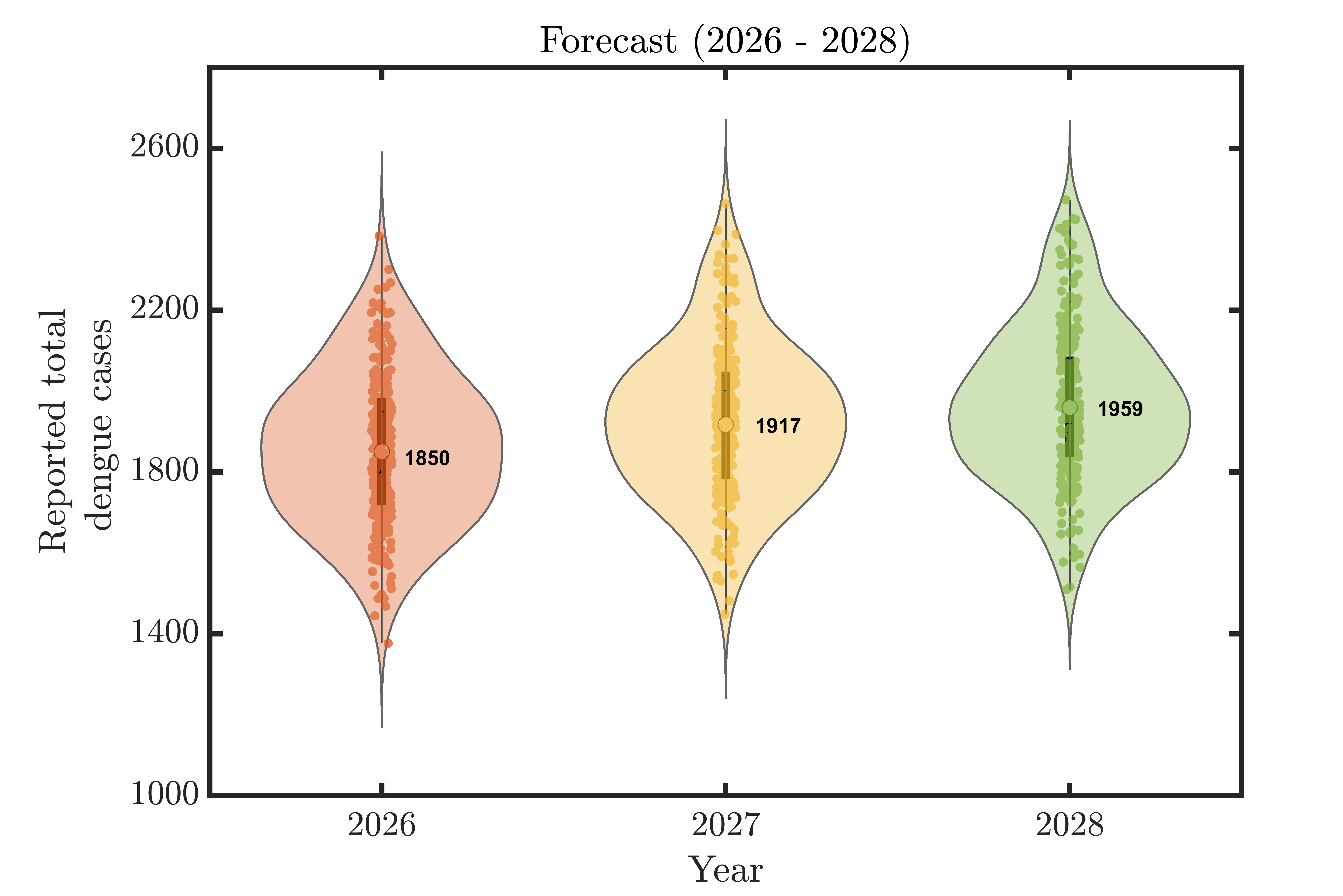} 
\caption{\textbf{Probabilistic prediction of annual dengue incidence in 2026–2028.} Three violin plots demonstrate the distribution of predicted numbers of reported dengue cases, where the colours correspond to 2026 (orange), 2027 (yellow), and 2028 (green). Single dots indicate predictions derived from different bootstrapped ensembles, demonstrating the variability of the ensemble itself. The middle point is the median estimate, while the wide horizontal line depicts the interquartile range. The entire width of each violin depicts the range between the 2.5 and 97.5 percentiles, which correspond to the 95 percent confidence intervals of the projection.}
\label{forecast}
\end{figure}

\subsection{Forecast of dengue incidence (2026--2028)}
Based on the parameter ensemble generated by the bootstrap method as discussed in Section~\ref{methods}, we forecast the dengue disease burden from the years 2026 to 2028. The probability distributions for annual dengue case reports are illustrated in Fig.~\ref{forecast}.

For Fig.~\ref{forecast}, the violin plots of the dengue cases in 2026, 2027, and 2028 are denoted using different colours (orange, yellow, and green, respectively). The points plotted in each violin denote the results of bootstrapping for each year. The point inside each violin represents the estimated median value, whereas the wider black line represents the interquartile range. The entire vertical span of the violin indicates the 2.5\%-97.5\% quantile range, which is equal to the 95\% confidence interval of the forecast distribution. The model estimates that the median annual cases would reach 1850 in 2026, 1917 in 2027, and 1959 in 2028, respectively, implying an increasing number of cases of dengue infection in the future. An increasing trend in the case numbers is seen in the forecasts made using the bootstrap technique; the medians and quantiles, in general, have shifted upwards over time. Specifically, the heavier right tail in 2028 implies a possible high-risk scenario.

Concerning the epidemiology of this situation, the gradual rise can be considered a continuation of the endemic nature of dengue infections in the AMC. Without any strong intervention from outside sources, the human-mosquito system will continue to maintain endemic infections through vector amplification during the seasons and through the constant supply of new susceptibles. This is how the dynamics of dengue infection typically operate within endemic situations, whereby after a brief drop, there is usually a resurgence because of the changing environmental and climatic conditions. The spread of the forecasted distributions in Fig.~\ref{forecast} shows that the variability of the future dynamics is largely determined by parameter uncertainties. The variations in the different bootstrap samples of the parameters used here are largely caused by the uncertainties of the parameters related to the transmission intensity and the symptomatic fraction.

In conclusion, Figure~\ref{forecast} clearly shows that the developed model~\eqref{eq:model}, together with the bootstrap framework, is very capable of providing a probabilistic description of the incidence of future dengue cases. As seen, there will be a continuous and somewhat increasing trend of disease burden in the next few years, highlighting the necessity of early interventions.
  
    \begin{figure}[h]
        \centering
        \includegraphics[width = \linewidth]{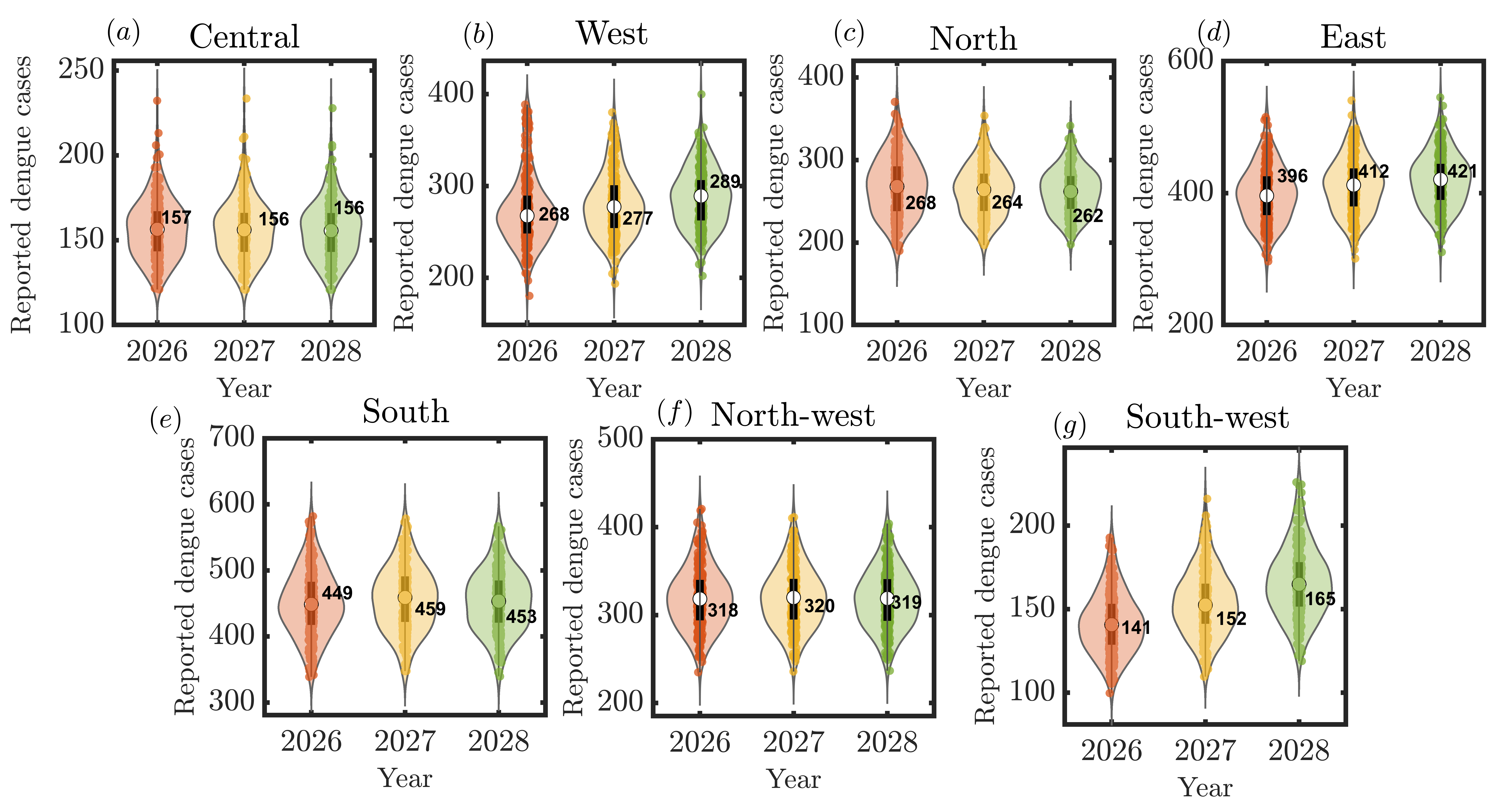} 
		\caption{\textbf{Forecasting of dengue incidence rate per region of AMC zones (2026--2028).} The figure illustrates the probability distribution of the total number of annual cases of dengue disease based on simulations of the M-SDT model~\eqref{eq:model}. The following six AMC zones have been considered: ($a$) Central Zone, ($b$) West Zone, ($c$) North Zone, ($d$) East Zone, ($e$) South Zone, ($f$) North-West Zone, and ($g$) South-West Zone.}
		\label{zonewise_forecast}
	\end{figure}

\subsubsection{Zone-wise forecasting of dengue incidence across AMC regions}
To account for the spatial heterogeneity in the dengue transmission at AMC, the aggregate forecast predictions are expanded into zone-wise projections through the seven municipal zones of the city. Figure~\ref{zonewise_forecast} shows the forecasted distributions of annual incidence rates of dengue infections for the period between 2026 and 2028, based on the bootstrap prediction from the calibrated model~\eqref{eq:model}. The parameter estimation is performed for each zone individually, using the procedures described in Section~\ref{methods}, and the fitted models along with their parameter distributions are included in the Supplementary Information. The data collected during the years 2020--2024 show high spatial variability, where different zones demonstrate different time dynamics, such as sudden outbreaks, partial decreases, and steady state.

While the Central zone (Fig.~\ref{zonewise_forecast}$(a)$), which witnessed relatively stable fluctuations in the past, still manages to retain its stable and low endemicity, it indicates that there are no powerful amplification systems in this area. Conversely, the West zone (Fig.~\ref{zonewise_forecast}$(b)$), which has been consistently transmitting the disease in recent times, indicates a steady increase in the projections, implying that transmission conditions remain favourable. The North zone (Fig.~\ref{zonewise_forecast}$(c)$) indicates that the region is now in a stabilized regime after being in a high burden zone for some time before, as indicated by the projected downward trend in case numbers. The East zone (Fig.~\ref{zonewise_forecast}$(d)$), which has been known to have high incidence for the longest period of time, continues to have elevated case numbers with a slight upward trend. Likewise, the South zone (Fig.~\ref{zonewise_forecast}$(e)$), which had very high incidences for the most part of the peak years, maintains a highly elevated projected incidence. The North-West zone (Fig.~\ref{zonewise_forecast}$(f)$), which was characterized by fluctuation in recent years, now settles down to a higher transmission zone, while the South-West zone (Fig.~\ref{zonewise_forecast}$(g)$), which has always had relatively low incidences, has a rising projected incidence.

The spread of the violin plots across all the zones indicates the presence of uncertainties related to estimation of parameters and stochastic fluctuations during the course of the transmission process. In conclusion, the forecast of the zones shows that the impact of previous epidemics is still being felt, with some zones showing hot spots while others show emerging hot spots.

\subsection{Impact of intervention strategies}
In order to explore the effectiveness of vector control in reducing dengue transmission, we analyze time-varying intervention processes in which the death rate of mosquitoes is altered within the M-SDT model paradigm. The mathematical derivation of such an intervention strategy is discussed in detail in Section~\ref{mechanism_intervention}. Two widely used interventions, i.e., fogging and residual spraying, have been included in our analysis framework \cite{senapati2019impact}.

There exist differences in their implementation procedures, where fogging is applied periodically in the short run, while spraying offers a more sustained approach that decays gradually. Evaluation of these approaches’ effectiveness is done through comparisons of predicted dengue cases in varying seasonal application scenarios for the forecasting period of 2026–2028.

\begin{table}[htbp]
\centering
\caption{Percentage reduction in total dengue cases under different intervention strategies.}
\label{tab:intervention_reduction}
\begin{tabular}{llccc}
\hline
\textbf{Intervention} & \textbf{Duration / Period} & \textbf{2026 (\%)} & \textbf{2027 (\%)} & \textbf{2028 (\%)} \\
\hline

\multicolumn{5}{c}{\textit{Fogging (four months)}} \\
June--Sept  & Weekly & 19.22 & 68.84 & 83.53 \\
July--Oct   & Weekly & 9.38  & 60.88 & 78.25 \\
\hline

\multicolumn{5}{c}{\textit{Fogging (three months)}} \\
June--Aug & Weekly & 18.51 & 62.83 & 77.88 \\
July--Sep & Weekly & 8.93  & 53.63 & 69.70 \\
Aug--Oct  & Weekly & 2.51  & 49.43 & 66.76 \\
\hline

\multicolumn{5}{c}{\textit{Spraying (six months)}} \\
June--Nov & Continuous & 22.94 & 79.43 & 93.13 \\
July--Dec & Continuous & 11.88 & 75.33 & 91.56 \\
Aug--Jan  & Continuous & 4.14  & 76.94 & 92.55 \\
\hline

\end{tabular}
\end{table}

\begin{figure}[htp]
\centering
\includegraphics[width = 1.0 \linewidth]{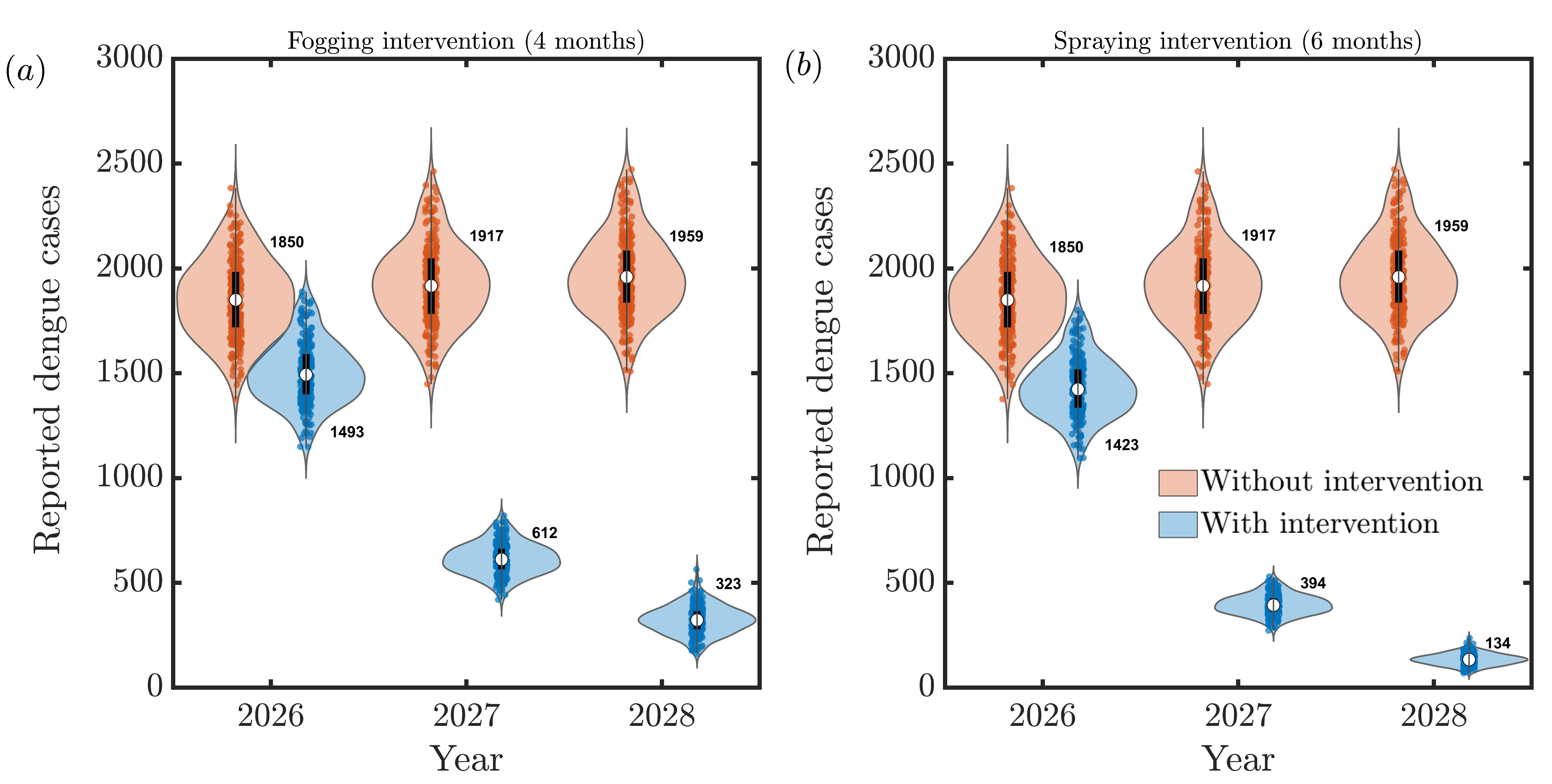} 
\caption{\textbf{Violin plots depicting the distribution of the number of yearly reported cases of dengue in 2026--2028 based on the bootstrapping results of the SEIAR-mosquito model~\eqref{eq:model}:} Red violin plots correspond to forecasted scenarios in the absence of interventions, whereas blue violin plots correspond to forecasted scenarios with interventions. In ($a$), fogging is performed for four months per year, whereas in ($b$), spraying is conducted for six months per year using a decay rate of $1/2$. The white points refer to the median, the vertical bars show the interquartile range, and the numbers refer to the median values of reported dengue cases.}
\label{intervention_png}
\end{figure}

\subsubsection{Reduction in dengue incidence under seasonal intervention strategies}
By building on the predictions made using the calibrated model \eqref{eq:model}, we evaluate the effect of implementing dynamic interventions by considering different periods during which the interventions are implemented. The percent reduction in the number of dengue cases is shown in Table \eqref{tab:intervention_reduction} for 2026–2028 with respect to different dynamic intervention plans depicted in Fig.~\ref{functions}$(a)$--$(b)$ (functional forms discussed in Section~\ref{mechanism_intervention}).

For the next strategy, we use a periodic impulse in form of fogging which aims at controlling the population of adult mosquitoes. Inspired by the vector-control process and the seasonality associated with the transmission of dengue fever, fogging takes place for three and four months each year. For the four-months intervention, the optimal period in terms of cases reduced is from June--September where the case reduction for the respective years of 2026, 2027, and 2028 is $19.22\%$, $68.84\%$, and $83.53\%$ respectively (Table \eqref{tab:intervention_reduction}; the distribution for the intervention cases reduction is provided in Fig.~\ref{intervention_png}$(a)$). Conversely, a late onset intervention period of July–October leads to relatively smaller decreases of ($9.38\%$, $60.88\%$, $78.25\%$), underlining the need for an earlier onset of the control measures in order to minimize the outbreak propagation. The same is also true for the three-month fogging period, although its impact is less effective since the duration of the intervention is shorter. Of the three options, the intervention period of June–August emerges as the most efficient one, with reductions of $18.51\%$, $62.83\%$, and $77.88\%$ for the three-year forecasts. Moving the onset of the intervention window towards later months of the transmission season ($July–September$ and $August–October$) reduces the effectiveness of the impulsive control measures, especially in 2026, when it drops to $2.51\%$.

On the other hand, spraying is formulated as a continuous method of intervention where the time decay function has a smooth trend (see Fig.~\ref{functions}$(b)$). Because of the longer lifespan of the spray intervention, spraying will be applied for half a year. In terms of reduction rates, the window between June and November performs well, with case reductions of $22.94\%$ in 2026, $79.43\%$ in 2027, and $93.13\%$ in 2028 (see Table~\eqref{tab:intervention_reduction}). Despite a delay in applying the intervention (e.g., July--December and August--January), reductions are still relatively high, at more than $90\%$ by 2028. This robustness results from the constant suppression of mosquitoes by the quadratic decay process associated with the spraying term $s(t)$ (see Fig.~\ref{functions}$(b)$). This constant process differs from that of fogging, where there is no continuity in the impact but rather localized events that take place at discrete times.

In summary, there are two major mechanistic lessons that can be learned from the comparison study. The first lesson is that impulsive measures, like fogging, are extremely dependent on the timing of their implementation and can gain much from an early implementation at the beginning of the season. The second lesson is that continuous measures, like spraying, have more resilience and efficacy since they have a lasting effect on the environment. The decreasing trends between 2026 and 2028 illustrate that repeated measures during each year build up over time, making the transmission cycle less effective.

\section{Discussion and Conclusion}\label{sec_discussion_conclusion}

The current paper provides a mechanistic analysis of the transmission dynamics of dengue fever in the Ahmedabad Municipal Corporation (AMC) using data and considering epidemiology, seasonality, and control measures. The analysis showed that the dengue transmission process in AMC is endemic and persistent rather than sporadic, driven by seasonal growth of mosquito population numbers. The predicted rise in the number of cases over 2026--2028 is indicative of an intrinsically dynamic system rather than a self-limiting one. Endemicity in other urban dengue systems has been driven by climatic variation and population dynamics~\cite{siraj2017temperature}. The key findings of the study are as follows

\begin{itemize}
\item The model shows a slow upward trend for the incidence of dengue, with the median number of cases per year increasing from roughly 1850 in 2026 to 1960 in 2028.
\item Since the reproduction number $R_0 > 1$, the disease remains endemic in AMC.
\item The sensitivity analysis highlights that the mosquito biting rate and mosquito death rate play crucial roles in the transmission process.
\item Measures like spraying can lower the incidence of dengue by over 90\%, but the impact of fogging heavily relies on its timing.
\end{itemize}

One important finding that emerges from the analysis in this regard is that of the hidden infection load. According to the bootstrap-based approach used for parameter estimation, biting rate is strictly regulated and there is high variability in the value of the symptomatic fraction. Estimates of the symptomatic fraction show that the presence of asymptomatic infections prevails overwhelmingly. As has been demonstrated in empirical studies before, the contribution of asymptomatic infections to the infection dynamics is highly significant. This makes it important to consider a surveillance framework that can take into account subclinical infections, too, in estimating disease load in AMC. Moreover, the spatiotemporal organization of dengue incidence also shows that the disease spread process is extremely heterogeneous across different AMC regions. The simultaneous presence of areas with high burden and emergent hotspots implies that the dynamics of dengue spread depend on local environmental and infrastructure factors, as opposed to citywide phenomena. This has been observed to maintain disease transmission due to local human-vector interactions~\cite{salje2017dengue}. This shows that interventions should be adaptable in space according to their risk maps, and resource allocation should be based on the changing nature of their risks. The mechanistic implementation of the intervention strategy sheds light on the feasibility of controlling dengue transmission. Impulsive fogging interventions are highly sensitive to seasonal timing, and earlier application increases their efficiency significantly. This finding aligns with previous studies suggesting that vector control interventions need to be synchronized with seasonal dynamics to have significant impacts~\cite{scott2009vector}. On the other hand, interventions such as spraying, which act continuously and persistently, show more resilience and prolonged mosquito population suppression. This indicates that temporary actions alone will not suffice; persistent, seasonally synchronized interventions are necessary for multiple-year stabilization of transmission dynamics.

Policy-wise, the results from the analysis have immediate significance regarding dengue control in AMC. With continued disease transmission and heterogeneity in space, in addition to delays in interventions, it is clear that reactive measures will not suffice. Instead, proactive, seasonally aligned, and spatially targeted interventions must be prioritized. Sustained vector control measures, supported by improved surveillance and data integration, are essential to reducing long-term disease burden and preventing recurrent outbreaks. However, the simulation scenarios can be further fine-tuned based on the requirements of the local policy makers.

Beyond these mechanistic findings, the present framework highlights the growing importance of integrating data-driven methodologies with epidemiological modelling. While compartmental models provide interpretability and biological consistency, the variability observed in parameter estimation and forecasts indicates that real-world transmission dynamics are influenced by factors that extend beyond the scope of purely mechanistic assumptions. However, the model has a few limitations. The model assumes a constant reporting rate and does not explicitly incorporate climate variability. Human mobility and fine-scale environmental heterogeneity are not explicitly modelled.  Incorporating high-dimensional data streams, such as climate variability, urban infrastructure, and human mobility, within hybrid modelling frameworks offers a pathway toward improving predictive accuracy. These integrated modelling approaches might account for dynamical components which elude conventional models~\cite{lourenco2014dengue}. Specifically, the combination of machine learning methods with mechanistic approaches could allow for dynamic model calibration and forecasting that is much more adaptive to changing conditions. Taken together, these factors suggest a need for a paradigm change in how dengue models can contribute to public health policy. More accurate estimation of asymptomatic transmission and the inclusion of spatial heterogeneity in decision making processes, along with adaptive data assimilation approaches for forecasting and intervention planning, are likely to become key areas of focus going forward.

In summary, dengue transmission in AMC can be described as a dynamically sustained and spatially structured phenomenon that demands coordinated interventions. The application of the methodology proposed in this work allows one to develop an appropriate approach for analyzing urban dengue transmission dynamics. This paper highlights that the integration of data and modelling techniques will constitute a key aspect in developing accurate forecasting tools.

 \begin{figure}[h]
        \centering
        \includegraphics[width = 1.0 \linewidth]{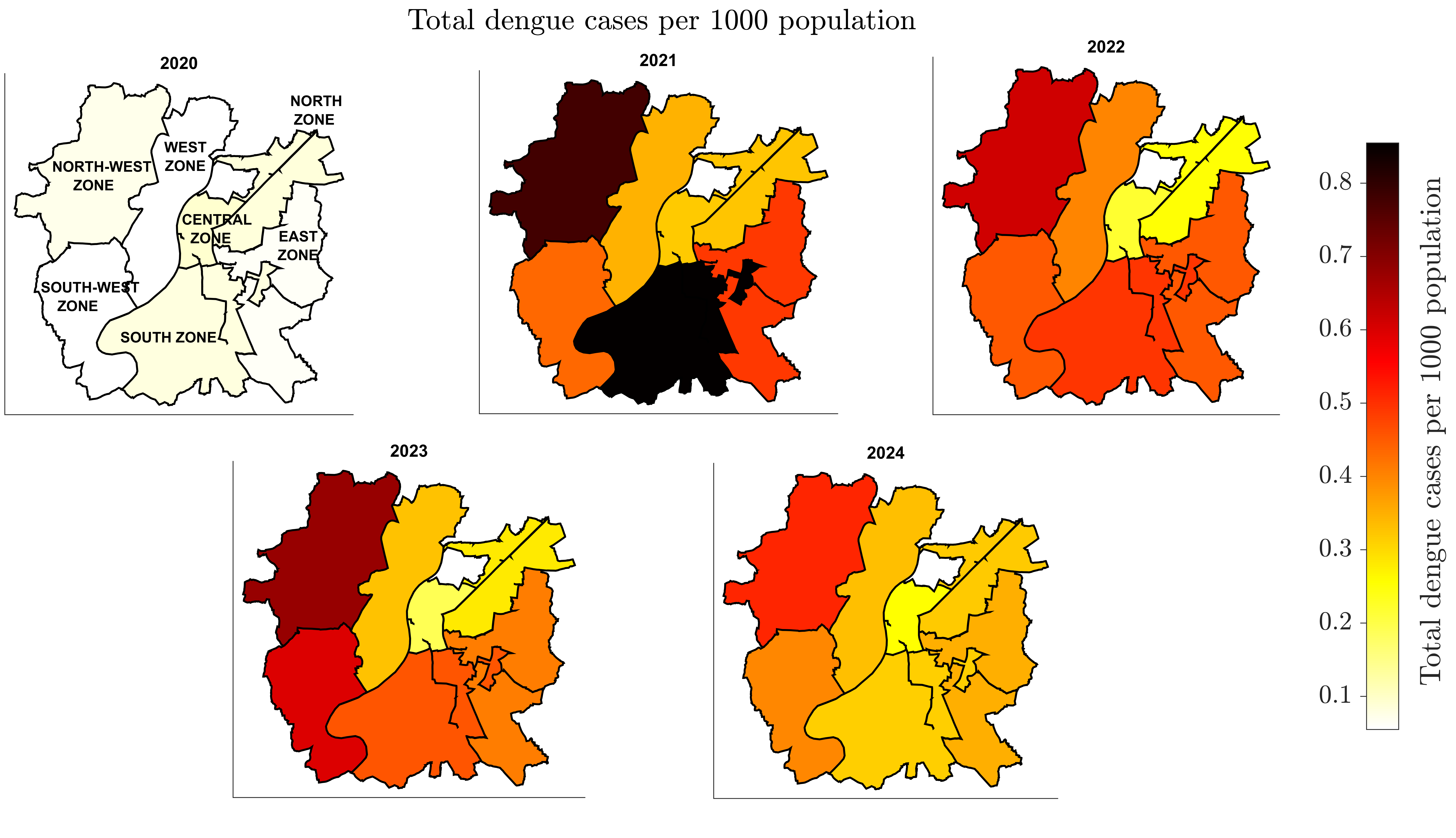} 
		\caption{{\bf Dengue cases per year, zoned-wise division of dengue patients in Ahmedabad city, from 2020 to 2024:} Annualized total dengue patients (including government and private hospitals' data) from 2020 to 2024 for AMC zones on a map basis, normalized by per 1,000 population, in which each figure is obtained by aggregating dengue cases for a particular year among all AMC zones. Color coding indicates higher total dengue patients load from lighter colors to darker colors, i.e., from less incidence of dengue to severe outbreak of dengue disease.}
		\label{fig:total}
	\end{figure}

\section{Materials and Methods} \label{sec_methods_materials}
This part describes the mathematical and computational methodologies used to generate the outcomes in the current study. On top of the existing modelling paradigm~\eqref{eq:model} introduced in earlier parts of this paper, the methodology section describes how the parameters are calibrated, uncertainty is quantified using bootstrap, and forecasts are generated. Moreover, we explain how time-varying functions $f(t)$ and $s(t)$ are implemented as part of the intervention strategies.

\subsection{Spatio-temporal distribution of dengue incidence in Ahmedabad (2020--2024)} \label{total_plot}

In order to understand the dynamics of dengue transmission through AMC, we conduct spatio-temporal analyses based on normalized incidence cases per 1,000 people in each AMC zone (Fig.~\ref{fig:total}) based on yearly incidence cases for each zone in AMC. Normalizing the dengue incidence rate helps compare the burden of the disease from one zone to another, irrespective of different population sizes of each zone, giving a better picture of transmission dynamics of the disease~\cite{roy2025spatio}. Figure~\ref{fig:total} shows the spatial distribution of total dengue incidence within seven zones in the AMC during the 2020-2024 period.

The level of dengue incidence was low in all the zones in 2020, with values falling within the lowest end of the color spectrum. Such a trend is expected due to the reduced cases of transmission during the time of the COVID-19 pandemic as a result of movement restrictions. There is a noticeable rise in the number of incidence cases from 2021. The year 2021 witnessed the largest outbreak of dengue virus cases since there was an obvious rise in the numbers for all zones, especially the Central and Southern zones. However, the East and Western zones had a relatively large increase in the number of cases. From 2022, the incidence was generally lower compared to 2021, but some zones, like the Western, Southern, and North-Western zones, have maintained their incidence rates. In 2023, there is even greater evidence of movement in terms of dengue cases in relation to the western and north-western sections of the city, making these areas the key hot spots. This suggests that there has been a geographical change in the distribution of transmission risks, possibly caused by urban growth, environmental factors, and unequal vector control measures. In 2024, there is a trend of a decreased number of cases occurring in all sections except for some selected ones.

In conclusion, the spatiotemporal dynamics show that there are periodic outbreaks and changing geographic locations of high transmission. The continuous high incidence in some areas emphasizes the requirement for strategic interventions. Further details of the government hospitals and private hospitals reporting of cases have been presented in the Supplementary information section, with spatiotemporal patterns evident even there.

\subsection{Parameter estimation, bootstrap resampling, and forecasting framework} \label{methods}

In constructing the procedure for parameter estimation and forecasting, the framework is developed as an inferential process based on the dynamical system described by equation \eqref{eq:model}. Initialization of the M-SDT model begins with extensive simulations starting from 1961 \cite{Gujarat1972} using the state vector

\[
X(t) = (S(t), E(t), I(t), A(t), R(t), M_S(t), M_I(t))^\top
\]
is developed until the point where a dynamically consistent situation is obtained. The final state of this evolution is employed as the starting point in future calibrations, guaranteeing the accuracy of the historical demography and epidemiology.

In the first stage, an equilibrium fitting is performed to estimate the parameters $(b_0, p, m)$ by aligning the model output with early incidence data. This step determines the mosquito-to-human ratio as $m = 3.098$, which is subsequently fixed to avoid parameter non-identifiability in the reduced estimation problem.

In the second stage, the remaining parameters $(b_0, p)$ are estimated by fitting the model to annual dengue incidence data from 2021--2024. The model-predicted annual cases are obtained by aggregating the daily incidence derived from influx to the symptomatic infected compartment. We consider a constant reporting rate ($\rho$) of 5\%. Specifically, the instantaneous reported incidence is given by $\mathcal{I}(t) = \rho \, p \, \sigma_E \, E(t)$, and the annual incidence is computed as
\[
\mu(t_y) = \int_{t_y}^{t_y+365} \mathcal{I}(t)\,dt,
\]
where $t_y$ denotes the starting time of year $y$. The parameter estimation problem is then formulated as a nonlinear least-squares minimization:
\[
\min_{b_0,\,p} \sum_{y=2021}^{2024} \left( C_y - \mu(t_y; b_0, p) \right)^2,
\]
where $C_y$ represents the observed annual cases.

To incorporate variability and overdispersion in reported data, we adopt a parametric bootstrap framework based on a negative binomial observation model. Given the model-predicted mean $\mu(t_y)$, synthetic observations are generated as
\[
C_y^{(k)} \sim \mathrm{NB}\left(r,\, p_y\right), \quad p_y = \frac{r}{r + \mu(t_y)},
\]
where $r=24$ is the dispersion parameter. This yields $\mathbb{E}[C_y^{(k)}] = \mu(t_y)$ and $\mathrm{Var}(C_y^{(k)}) = \mu(t_y) + \mu(t_y)^2/r$, thereby capturing overdispersed count variability.

For each bootstrap realization $k = 1, \dots, M = 250$, the parameter pair $(b_0^{(k)}, p^{(k)})$ is re-estimated by solving the same least-squares problem using the synthetic dataset $\{C_y^{(k)}\}$. This produces an ensemble of parameter estimates 
\[
\Theta = \{(b_0^{(k)}, p^{(k)})\}_{k=1}^M,
\]
along with corresponding model trajectories $X^{(k)}(t)$.

The forecasting procedure is then carried out by propagating each parameter realization forward in time. For each $(b_0^{(k)}, p^{(k)}) \in \Theta$, the system~\eqref{eq:model} is integrated each year over the interval corresponding to 2026--2028, and annual incidence is computed as
\[
Y_y^{(k)} = \int_{t_y}^{t_y+365} \rho \, p^{(k)} \, \sigma_E \, E^{(k)}(t)\,dt.
\]
This yields, for each year $y$, a distribution $\{Y_y^{(k)}\}_{k=1}^M$ of predicted cases~\cite{chowell2017fitting}.

To characterize these distributions, we employ kernel density estimation (KDE) to obtain a smooth approximation of the empirical density:
\[
\hat{f}_y(x) = \frac{1}{Mh} \sum_{k=1}^M K\left(\frac{x - Y_y^{(k)}}{h}\right),
\]
where $K(\cdot)$ is a Gaussian kernel and $h$ is the bandwidth. Using these densities, violin plots can be generated, which will enable the visualization of the whole distribution, central tendency, and dispersion of the forecast. The above method guarantees that uncertainties due to parameter estimation are accounted for in the non-linear model dynamics leading to probability forecasts.

\begin{figure}[ht]
\centering
\includegraphics[width = 1.0 \linewidth]{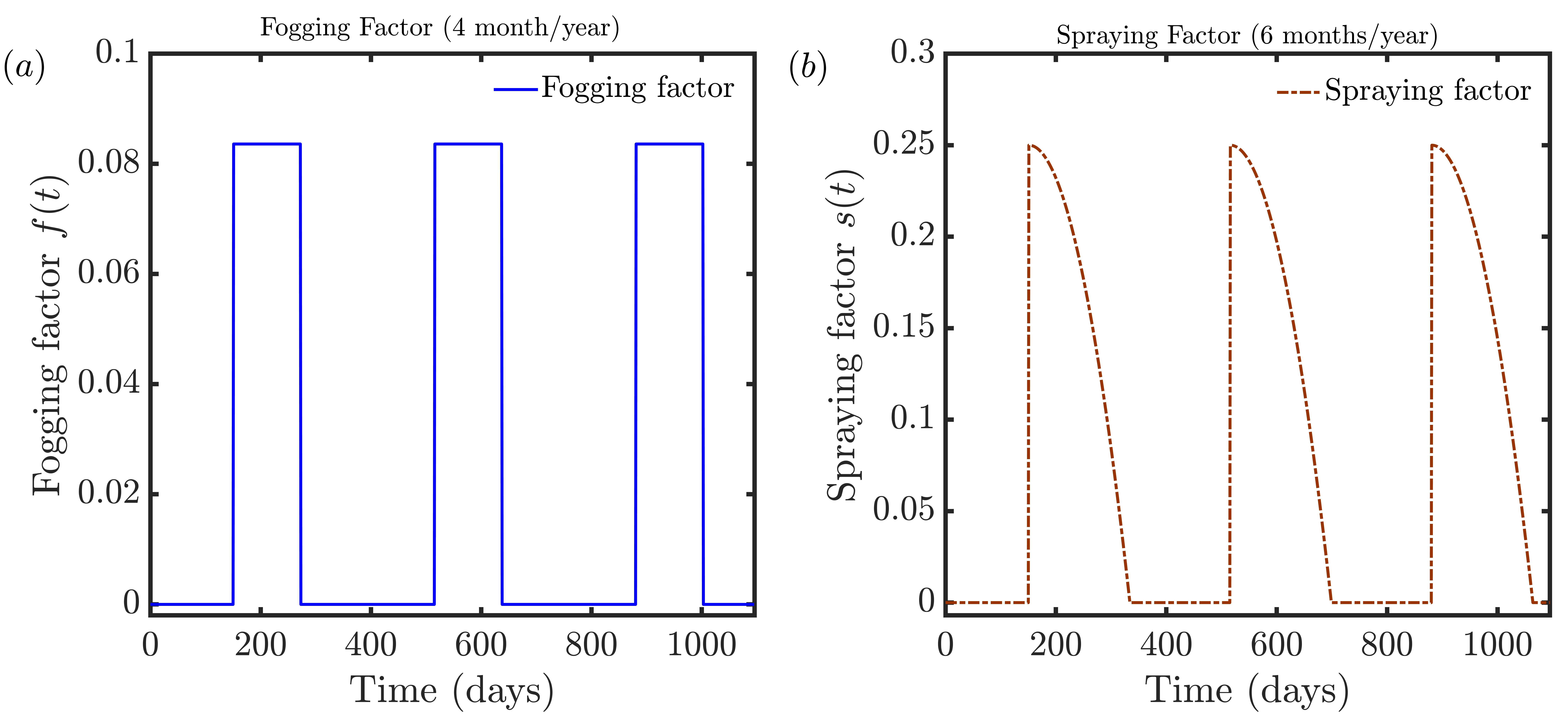} 
\caption{\textbf{Time-wise deployment of vector control measures during three consecutive years.} 
($a$) Time-variant fogging rate $f(t)$, conducted for four months each year (June–September), presented by a periodic stepwise function reflecting the timing of post-monsoon vector control campaigns.
($b$) Time-variant spraying rate $s(t)$, carried out for six months each year (June–November), showing a decay curve depicting the diminishing effects of spray efficacy after each spraying period.
These control measures were formulated consistently with the principles outlined in Sections~3.5 and~5.3.}
\label{functions}
\end{figure}
\subsection{Adult mosquito control strategies}
Dengue prevention can be attained using either larval or adult mosquito management techniques. But in areas where the disease has been actively spreading, adult mosquito management is more pertinent because the spread of the disease is brought about by the biting behavior of the adults. Hence, adult vector management becomes extremely important during outbreaks and after monsoons.

This paper analyzes two widely used adult mosquito control programs for their effect on the dynamics of dengue transmission in AMC. The two control methods differ in their operating principle, efficiency, and environmental longevity. The description of the two methods is given below.

\noindent
\textbf{C1. Fogging (Spraying of insecticides in ultra-low volume):}
Fogging consists of the spraying of insecticides in ultra-low volumes using ULV cold fogging sprayers. The sprayed insecticides form tiny droplets that stay suspended in the air temporarily but have an impressive capability of killing large numbers of adult mosquitoes. Given its prompt results, fogging is often used during times of peak disease transmission. Nonetheless, it is relatively short-lived since the effects of the sprayed insecticides are temporary and wear off within a day.

\medskip

\noindent
\textbf{C2. Residual spraying (Insecticide surface treatments):}  
In residual spraying, insecticides are sprayed on indoor and outdoor surfaces including walls, curtains, and water tanks used for storing water. This technique helps in decreasing mosquito survival rates by minimizing interactions between mosquitoes and humans and increasing mosquito deaths after coming into contact with insecticidal surfaces. The difference between residual spraying and fogging lies in the fact that residual spraying has long-lasting effects in the environment, lasting up to a few months.

\medskip

The implementation of such strategies results in excess mortality among mosquitoes. Each strategy possesses its own unique features in terms of efficiency and sustainability; therefore, the impact of each of them on the level of mortality is inherently time-dependent. For example, the fogging strategy results in a transient surge in mortality, while the spraying strategy causes a more sustainable mortality rate that is slowly decreasing.

In order to take into account these factors in our model, it is necessary to include control functions with respect to time, which will affect the basic death rate of mosquitoes.

\subsubsection{Intervention strategies} \label{mechanism_intervention}

To incorporate external control measures into the M-SDT model~\eqref{eq:model}, we modify the mosquito mortality terms through time-dependent functions. In the baseline formulation, mosquito dynamics include the natural mortality rate $\mu_M$ through the terms $-\mu_M M_S$ and $-\mu_M M_I$. Under intervention, this mortality rate is augmented by additional time-dependent contributions, resulting in modified decay terms in the mosquito compartments.

Two intervention functions are introduced, denoted by $f(t)$ and $s(t)$, representing fogging and spraying, respectively. 

\subsubsection{Fogging intervention} \label{fogging_analysis}
The intervention strategies are defined on a daily timescale and incorporated independently into the mosquito equations, leading to distinct modified dynamical systems.

Fogging is modelled as a periodic, seasonally confined intervention that induces a discrete increase in mosquito mortality over specified time intervals. The mosquito dynamics under fogging are given by
\begin{equation}
\begin{aligned}
\frac{dM_S}{dt} &= \mu_M N_M - \lambda_M M_S - \left(\mu_M + \,f(t)\right) M_S,\\
\frac{dM_I}{dt} &= \lambda_M M_S - \left(\mu_M + \,f(t)\right) M_I,
\end{aligned}
\label{fogging_fact}
\end{equation}
where $f(t)$ is the fogging intervention function.

The function $f(t)$ in Eq.~\eqref{fogging_fact} is defined on a daily timescale and constructed using a periodic mapping $d = t \bmod 365$ to represent annual recurrence. Let $d_s$ and $d_e$ denote the start and end of the fogging window, and let $E$ be the total number of fogging events within this period. The duration of the intervention window is $D$.

The fogging function is defined as
\[
f(t) =
\begin{cases}
\displaystyle \frac{0.6E}{D}, & d \in [d_s, d_e] \quad (\text{or periodic equivalent}), \\
0, & \text{otherwise}.
\end{cases}
\]

The multiplicative factor $0.6$ in our consideration represents a $60\%$ increase in mosquito mortality during active fogging periods. This formulation approximates repeated discrete fogging events (e.g., weekly applications) as a temporally averaged increase in mortality. The resulting function exhibits a piecewise-constant structure, corresponding to periodic activation of fogging within each seasonal cycle (Fig.~\ref{functions}$(a)$).

\subsubsection{Spraying intervention} \label{spray_analysis}
Residual spraying is modelled as a continuous intervention with a time-dependent intensity that gradually decreases following application. The mosquito dynamics under spraying are described by
\begin{equation}
\begin{aligned}
\frac{dM_S}{dt} &= \mu_M N_M - \lambda_M M_S - \left(\mu_M + s(t)\right) M_S,\\
\frac{dM_I}{dt} &= \lambda_M M_S - \left(\mu_M + s(t)\right) M_I,
\end{aligned}
\label{spraying_fact}
\end{equation}
where $s(t)$ denotes the spraying intervention function.

The function $s(t)$ in Eq.~\ref{spraying_fact} is defined over a seasonal window using a quadratic decay profile. Let $d = t \bmod 365$ denote the day within a yearly cycle, and let $d_s$ and $d_e$ represent the start and end of the spraying period. The duration of the intervention is denoted by $D$, and the elapsed time since the start of spraying is $\tau$.

The spraying function is defined as
\[
s(t) =
\begin{cases}
\displaystyle 0.25\left(1 - \left(\frac{\tau}{D}\right)^2\right), & d \in [d_s, d_e], \\
0, & \text{otherwise},
\end{cases}
\]
with
\[
\tau = d - d_s \quad (\text{or periodic equivalent}).
\]

The coefficient $0.25$ represents the maximum enhancement of mosquito mortality at the onset of spraying. The quadratic decay captures the gradual loss of effectiveness due to environmental degradation of insecticides. Consequently, $s(t)$ exhibits a smooth, concave temporal profile within each intervention window (Fig.~\ref{functions}$(b)$), in contrast to the step-like structure of the fogging function.

The intervention-modified systems defined above are integrated into the same computational framework described in Section~\ref{methods}, enabling the evaluation of the impact of each control strategy under consistent dynamical and statistical assumptions.

\subsection{Stability analysis of the SEIAR--mosquito model}

We analyse the qualitative behaviour of system~\eqref{eq:model}, including positivity, boundedness, the disease-free equilibrium (DFE), and the basic reproduction number $R_0$, which governs the threshold dynamics of the system~\cite{dank2014estimating}. For analytical tractability, the seasonal forcing term is omitted ($a_b = 0$), yielding an autonomous system.

\paragraph{Positivity and boundedness.}
Let $N_H = S + E + I + A + R$ denote the total human population. Summing the human equations gives
\begin{equation}
\frac{dN_H}{dt} = \Pi_H - \mu_H N_H - \xi I.
\end{equation}
Since $\xi I \geq 0$, we obtain
\begin{equation}
\frac{dN_H}{dt} \leq \Pi_H - \mu_H N_H,
\end{equation}
which implies
\begin{equation}
N_H(t) \leq \frac{\Pi_H}{\mu_H}, \quad \forall t \geq 0.
\end{equation}

Similarly, defining $N_M = M_S + M_I$, we obtain
\begin{equation}
\frac{dN_M}{dt} = 0,
\end{equation}
so that $N_M$ remains constant. Hence, the system evolves in the positively invariant region
\begin{equation}
\Omega = \left\{ (S,E,I,A,R,M_S,M_I) \in \mathbb{R}_+^7 : N_H \leq \frac{\Pi_H}{\mu_H}, \; N_M = \text{constant} \right\}.
\end{equation}

\paragraph{Disease-free equilibrium.}
In the absence of infection ($E = I = A = M_I = 0$), we obtain
\begin{equation}
S^* = \frac{\Pi_H}{\mu_H}, \qquad R^* = 0, \qquad M_S^* = N_M,
\end{equation}
and the disease-free equilibrium is
\begin{equation}
E_0 = \left( \frac{\Pi_H}{\mu_H},\, 0,\, 0,\, 0,\, 0,\, N_M,\, 0 \right).
\end{equation}

\paragraph{Basic reproduction number.}
To compute the basic reproduction number, we use the next-generation matrix approach. The infected subsystem is defined as
\[
X = (E, I, A, M_I)^\top.
\]
The dynamics of these compartments can be written as
\[
\frac{dX}{dt} = F(X) - V(X),
\]
where $F(X)$ represents the rate of new infections and $V(X)$ represents transitions between compartments and removals.

Linearising the system at the disease-free equilibrium $E_0$, we obtain the Jacobian matrices
\begin{equation}
F =
\begin{pmatrix}
0 & 0 & 0 & \dfrac{b_0 \beta_{HM} S^*}{N_H} \\
0 & 0 & 0 & 0 \\
0 & 0 & 0 & 0 \\
0 & \dfrac{b_0 \beta_{MH} M_S^*}{N_H} & \dfrac{b_0 \beta_{MH} \theta M_S^*}{N_H} & 0
\end{pmatrix},
\end{equation}

\begin{equation}
V =
\begin{pmatrix}
\sigma + \mu_H & 0 & 0 & 0 \\
-p\sigma & \gamma_I + \mu_H + \xi & 0 & 0 \\
-(1-p)\sigma & 0 & \gamma_A + \mu_H & 0 \\
0 & 0 & 0 & \mu_M
\end{pmatrix}.
\end{equation}

The next-generation matrix is then defined as
\begin{equation}
K = FV^{-1}.
\end{equation}
The basic reproduction number is given by the spectral radius of this matrix:
\begin{equation}
R_0 = \rho(K) = \rho(FV^{-1}).
\end{equation}

Carrying out the matrix inversion and multiplication yields the closed-form expression
\begin{equation}
R_0 =
\sqrt{
\frac{b_0^2 \, \beta_{HM} \, \beta_{MH} \, m}
{\mu_M}
\left(
\frac{p}{\gamma_I + \mu_H + \xi}
+
\frac{(1-p)\theta}{\gamma_A + \mu_H}
\right)
},
\end{equation}
where all parameters are defined in Table~\ref{table1}. This expression captures the combined contribution of symptomatic and asymptomatic transmission pathways mediated through the mosquito population.

\paragraph{Local stability of the disease-free equilibrium.}
The disease-free equilibrium $E_0$ is locally asymptotically stable if $R_0 < 1$ and unstable if $R_0 > 1$.

\paragraph{Effect of intervention on transmission dynamics.}
In the presence of control measures, the mosquito mortality rate is modified as $\mu_M \rightarrow \mu_M + u(t)$, where $u(t)$ represents intervention. Consequently,
\begin{equation}
R_0(t) \propto \frac{1}{\mu_M + u(t)},
\end{equation}
indicating that vector control reduces transmission by lowering the effective reproduction number.

\subsection{Basic reproduction number for AMC and its zones}

The basic reproduction number \(R_0\) serves as a threshold quantity that determines whether dengue transmission can invade and persist in a population. It is defined as the expected number of secondary infections generated by a single infected individual introduced into a fully susceptible population. For the present SEIAR--vector framework, \(R_0\) is obtained from the next-generation matrix derived in the stability analysis, and its numerical value is evaluated by substituting the calibrated parameters (Table~\ref{table1}).

Using the estimated parameters, the overall reproduction number for AMC is found to be
\[
R_0^{\text{AMC}} \approx 1.7929,
\]
which clearly exceeds unity, indicating that dengue transmission is endemic at the city scale.

To represent the heterogeneity present within the area, zone-wise values for \(R_0\) have been calculated using a bootstrap analysis method, providing 95\% confidence intervals (CI), as shown in Table~\ref{tab:R0_zones}.

\begin{table}[h!]
\centering
\caption{Zone-wise basic reproduction number \(R_0\) with 95\% confidence intervals.}
\label{tab:R0_zones}
\begin{tabular}{lcc}
\hline
\textbf{Zone} & \textbf{Mean \(R_0\)} & \textbf{95\% CI} \\
\hline
Central      & 2.2128 & [1.6695, 3.2110] \\
West         & 1.7473 & [1.5218, 1.9790] \\
North        & 1.6354 & [1.4941, 1.7499] \\
East         & 1.7966 & [1.5181, 2.0675] \\
South        & 2.0049 & [1.6198, 2.3803] \\
North-West   & 2.0562 & [1.6699, 2.8459] \\
South-West   & 1.9005 & [1.8126, 2.0006] \\
\hline
\end{tabular}
\end{table}

As per the theory, the epidemic threshold criterion states that:
\[
\begin{cases}
R_0 < 1 \quad &\Leftrightarrow \quad \text{disease-free equilibrium stable}, \\
R_0 > 1 \quad &\Leftrightarrow \quad \text{presence of endemic equilibrium}.
\end{cases}
\]

Values of \(R_0\) obtained through the bootstrap method satisfy \(R_0 > 1\) in all cases. Hence, every zone is in the state of endemic disease transmission within AMC. Nonetheless, the value of \(R_0\) gives us some additional insight into outbreak potential and transmission intensity:

\begin{itemize}
\item \textbf{Regimes with high transmission} (\(R_0 > 2\)): The Central, South, and North-West areas have relatively high values of \(R_0\). These zones represent a high transmission intensity regime where small changes in vector population and contacts result in big epidemics.

\item \textbf{Regimes with moderate transmission} (\(1.7 < R_0 \leq 2\)): The East and South-West zones demonstrate intermediate levels of transmission intensity; therefore, there is a possibility for an outbreak, but it is seasonally forced.

\item \textbf{Regimes with relatively controlled transmission} (\(1 < R_0 \leq 1.7\)): The West and North zones demonstrate relatively low levels of reproduction numbers.
\end{itemize}

The confidence intervals support this further. It is important to note that all lower bounds are greater than one, signifying that dengue transmission is endemic within each region. Moreover, the wider intervals seen in high-burden zones, such as Central and North-West, signify increased sensitivity, implying more variability in the response and the likelihood of experiencing more pronounced outbreak oscillations. From an epidemiological perspective, the results demonstrate that the underlying factor for differences in dengue disease burdens in AMC zones is based on variations in reproduction numbers. High-\(R_0\) zones not only facilitate the spread of infections better but may also act as sources for infections spilling over into neighboring zones.






\section*{Supplementary Information}
\section*{AMC zone-wise calculation for model fitting and forecasting}
In the main text, the model defined in Eq.~(3) is developed and calibrated using city-wide data for AMC. Correspondingly, we use the observed data for the total population of AMC to fit the model. Using bootstrap, we further obtain the forecasting and intervention strategies implemented globally to overcome the disease outbreak. Thus, here we apply the same technique to outline the path taken by the dengue epidemic in different regions, which is assessed using the seven zones of municipal corporations in Ahmedabad. 

\begin{figure}[ht]
	\centering
	\includegraphics[width = 1.0 \linewidth]{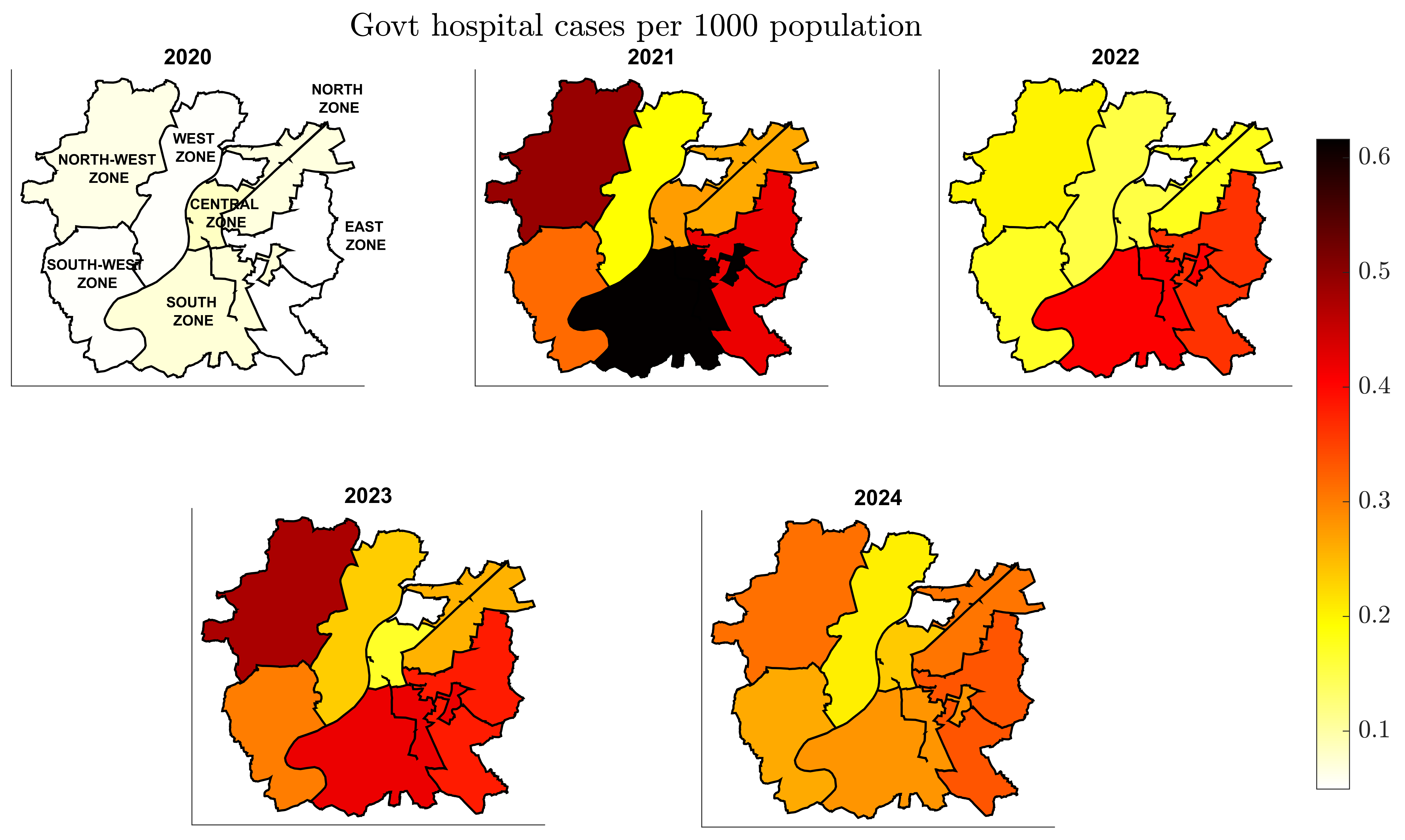} 
	\caption{\textbf{Spatiotemporal pattern of dengue incidence in government hospitals.} Spatiotemporal pattern of dengue incidence (per 1,000 population) for the AMC zones through reported cases at government hospitals from 2020 to 2024. These patterns show that there has been the development of a spatial cluster formation, with an increased incidence rate observed throughout the city in 2021.}
	\label{govt_hospital}
\end{figure}

\subsection*{Spatio-temporal heterogeneity of dengue incidence across AMC zones}

The distribution of dengue incidence per 1,000 inhabitants by zones indicates a highly heterogeneous and constantly changing transmission environment within AMC. Instead of being distributed uniformly in the entire city, dengue is distributed in a heterogeneous fashion in that particular city such that some areas may serve as reservoirs for the virus.

Figure \ref{govt_hospital} illustrates the pattern of dengue occurrence as reported in governmental hospitals. One remarkable aspect is the sudden surge witnessed in 2021, marking the shift from the lower endemic state in 2020 to the high-transmission period. The rise is not evenly distributed across the system but rather disproportionately affects the southern and eastern sectors of the city. The reason for this uneven pattern may be due to the presence of high population density, slums, or other such areas where the vector breeds. After 2021, there is no return to the previous stable condition of a lower incidence rate. The ongoing trend of high incidences in some zones, especially in the southern corridor region, during 2022 and 2023 points toward the creation of transmission reservoirs in those areas. However, there appears to be a progressive rise in cases from the western and north-western regions, marking the movement of the epidemic border. The spatial changes in disease dynamics have important implications because it shows that dengue virus transmission in AMC has a migratory pattern, which is a result of favorable environmental conditions and vector presence. While by 2024, there is an overall decrease in cases, the spatial effect of previous outbreaks still persists. Zones where there was once high transmission activity did not revert entirely to their baseline status, showing that the virus transmission process was not interrupted completely.

\begin{figure}[ht]
	\centering
	\includegraphics[width = 1.0 \linewidth]{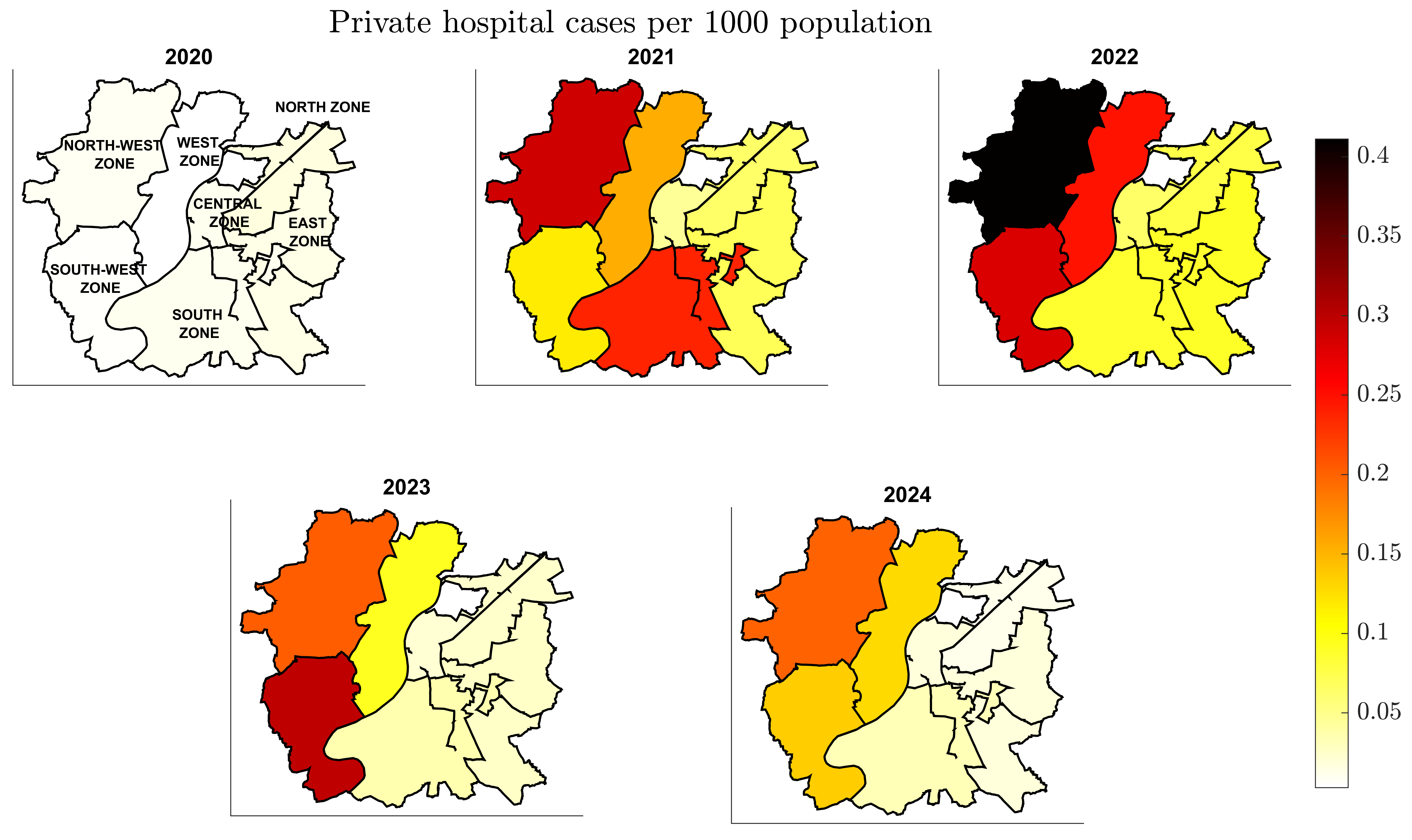} 
	\caption{\textbf{Spatio-temporal distribution of dengue incidence at private hospitals.} Spatio-temporal distribution of dengue incidence (per 1,000 individuals) for the AMC zones based on incidence data from private hospitals in the period 2020-2024. This reveals a spatiotemporally variable pattern of outbreaks, with distinct spikes for each zone as well as a relatively quick drop-off of incidence.}
	\label{private_hospital}
\end{figure}

However, Fig.~\ref{private_hospital} presents an entirely different but complementary insight into dengue incidence from the perspective of private healthcare information. Although the 2021 epidemic can still be clearly traced, the patterns of the disease distribution tend to be more fragmented and characterized by sharp contrasts between various zones. Specifically, the north-west zone becomes one of the leading zones with higher incidence rates than in other zones in 2022. Furthermore, there is no consistency between private and governmental hospitals' records concerning the number of cases since the former appear to be more susceptible to socioeconomic differences and healthcare access. In addition, the difference in terms of the speed of incidence contraction is notable since government hospitals recorded gradual contraction across zones from 2022 to 2024, while the private hospital cases experienced a steeper contraction rate after 2023. This difference might be attributed to a range of factors, such as differences in the way cases are reported, differences in the utilization of health services, and the likely changes in the way cases were detected during the post-outbreak period. The comparison between government hospitals' and private hospitals' databases helps us understand the dynamics behind transmission and reporting. The data from the government hospitals seem to include the wider epidemiology of the disease, while the data from the private hospitals seem to emphasize acute outbreaks.

These variations have immense importance from a public health perspective. First, the presence of zones having high incidences of dengue within government records implies a structural weakness such as poor sanitation, water management, and housing problems that need immediate attention. Second, the presence of local peak zones in the private data suggests inconsistency in outbreak detection and response within the city, which might lead to delayed interventions in areas that need it most. It is worth noting that the spatio-temporal variations seen in the data form the foundation of our model, since their nature suggests the presence of spatial and temporal heterogeneity within the system. The existence of mobile hotspots, the asynchronous reduction in incidence rates across zones, and the differences in healthcare systems within AMC imply that there are considerable spatio-temporal variations in dengue transmission.

All in all, it is shown that there is an intricate relationship between the process and the urban form concerning the incidence of dengue in AMC. With hot spots, transmission fronts, and reportage dependent on access to health care facilities, it becomes imperative to have interventions in place that target particular areas at certain times of the year. For zone-wise modelling, dengue cases are considered from government and private hospitals.

\section*{Zone-wise model calibration and bootstrap-based parameter estimation}
In order to conduct an analysis of the metropolitan area at the level of the individual zones within the city, as is discussed in detail in the main text (Section~(5.2)), the models are independently calibrated using data corresponding to each of the seven zones defined by the AMC.

One of the main difficulties in this case is the lack of information regarding the historical data for population distribution in zones for the base year. For this purpose, we assume the baseline population for each zone according to the 2011 census data. Let $N_0^{(j)}$ be the population in zone $j$ at time 2011 (According to Ahmedabad Municipal Corporation (AMC), as per 2011 Census of India, Zonal Demographic Data). Given that the net increase in population is driven by birth rates minus natural deaths, the population will change according to

\begin{equation}
	N_t^{(j)} = N_0^{(j)} \, \exp\big((b_H - \mu_H)\, t\big),
	\label{eq:zone_population}
\end{equation}

where $b_H = 7.92 \times 10^{-5}\ \text{day}^{-1}$ and $\mu_H = 3.0 \times 10^{-5}\ \text{day}^{-1}$ denote the per capita birth and death rates, respectively, taken from Table~(1). This equation guarantees consistency with the demographic parameterizations of the original model while allowing for population scaling on a zone basis.

Based on the estimated population dynamics from above, M-SDT models (see Eq.~(2) in the main text) are independently fitted to annual incidence rates of dengue fever cases within each area for 2021--2024. Similarly to the analysis at the city level, the value of the $m$ parameter, the ratio between mosquitoes and humans, is first estimated based on equilibrium conditions. Then, by fixing $m$, the transmission parameters are estimated via nonlinear least-squares regression, that is, the biting rate $b_0$ and the proportion of symptomatic cases $p$. To include randomness and noise in the process, a bootstrap procedure similar to the one in Section~(5.2) of the main text is utilized. That is, negative binomial synthetic data are produced based on fitted incidence levels, and the parameter values are recalculated.

This process produces a set of parameters $(b_0, p)$ per zone and associated trajectory fits. Such distributions generate confidence intervals on a per-zone basis and allow uncertainty to be calculated in the transmission dynamics. More importantly, this method of spatial calibration accounts for differences in the transmission rate, reporting, and sustainability of the disease outbreak in various zones of the city. The results of model fitting and associated parameter distributions per zone are displayed in the graphs that follow.

\begin{figure}[ht]
	\centering
	\includegraphics[width = 1.0 \linewidth]{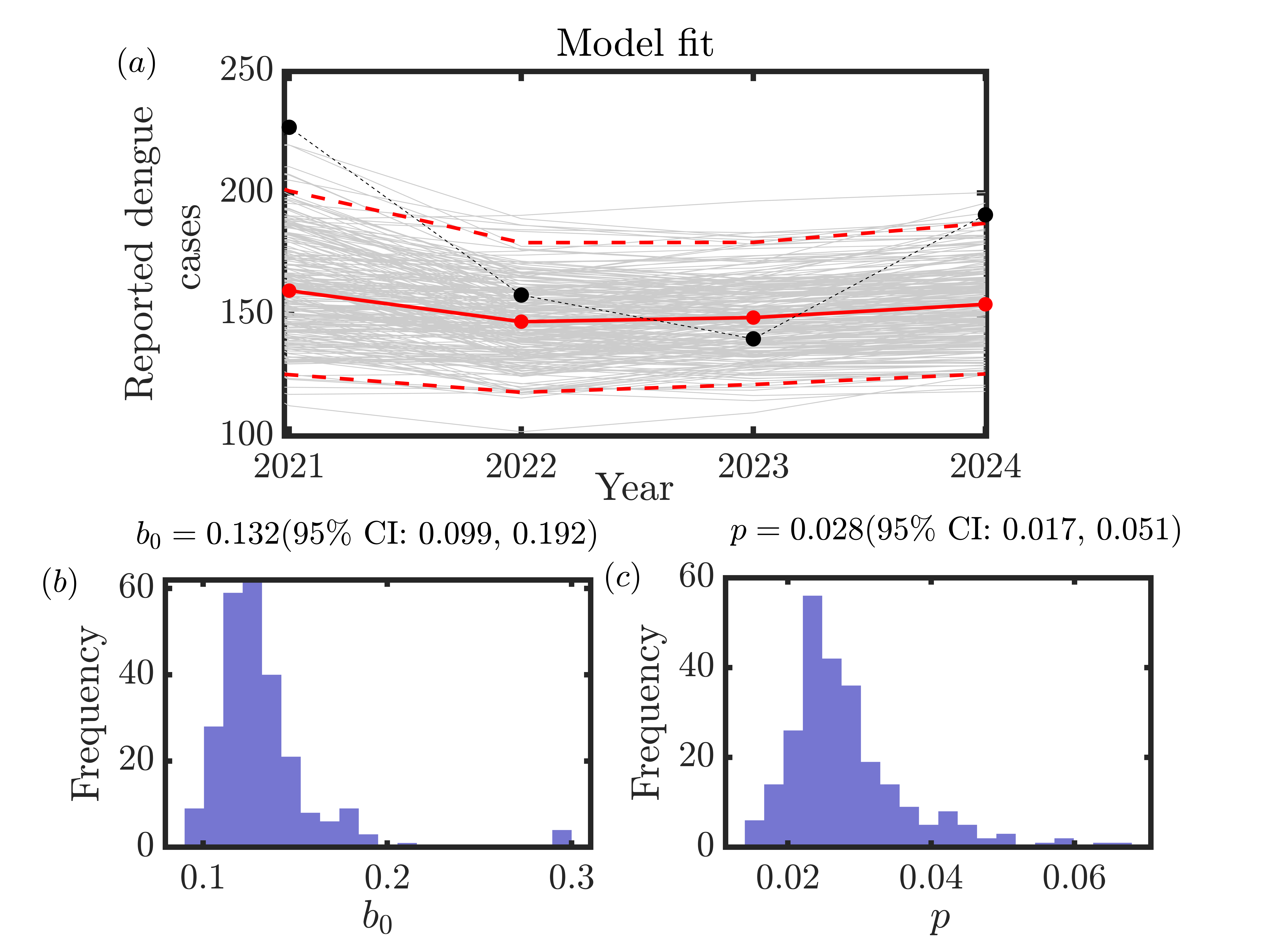} 
	\caption{\textbf{Central zone.} ($a$) Best-fit curve of reported dengue cases in the years 2021–2024, where black circles represent the reported data points, the red line shows the best-fit median curve, and gray curves represent the bootstrapped realizations. The dotted red lines show the 95\% confidence intervals for the best-fit curve. ($b$) Bootstrapped distribution of the baseline biting rate parameter \( b_0 \), with an estimate of its mean value and 95\% confidence intervals. ($c$) Bootstrapped distribution of the symptomatic fraction parameter \( p \).}
	\label{central_zone}
\end{figure}

\subsection*{Zone-wise model fitting, parameter estimation, and spatial heterogeneity}

\begin{figure}[ht]
	\centering
	\includegraphics[width = 1.0 \linewidth]{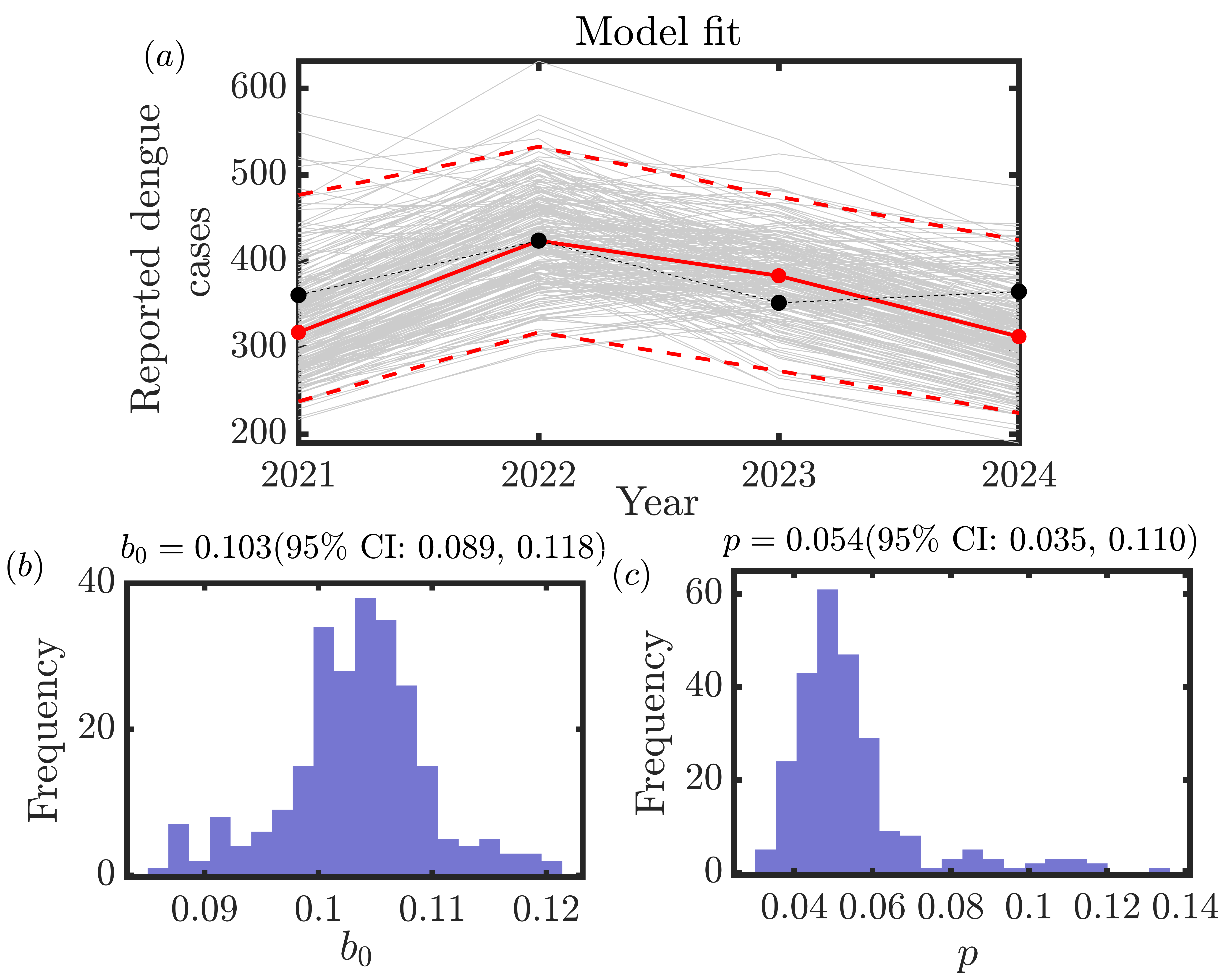} 
	\caption{\textbf{West zone.} $(a)$ Fit model depicting the evolution of the number of dengue cases over time, where bootstrap samples reflect the variability around the median path. $(b)$ Frequency histogram of the estimated baseline biting rate $b_0$, revealing a narrow distribution around the moderate parameter value. $(c)$ The distribution of the proportion symptomatic $p$, which is more dispersed and skewed to the right.}
	\label{west_zone}
\end{figure}

A detailed zone-wise calibration of the M-SDT model Eq.~(3) from the main text is presented here, which extends the city-wise calibration exercise to seven zones within Ahmedabad city. The associated fitted model and parameter distribution via bootstrapping for each zone are illustrated in Fig.~\ref{central_zone} through \ref{south_west}. There is significant spatial variability noted in the dynamics and structure of the disease in different zones.

\subsubsection*{Central zone}
Fitted to the data of the Central zone (Figure \ref{central_zone}$(a)$), the model reveals its consistent capacity to reconstruct the observed dengue incidence in the period from 2021 to 2024. Bootstrap runs remain bounded in a confidence envelope, meaning that the selected parameter values result in stable dynamics. The median of the ensemble (solid red line) shows that the decreasing tendency of incidence between 2021 and 2023 ends in slight growth in 2024. Thus, the dynamics of dengue incidence in this zone are rather regulated than amplified. In addition, such a narrow confidence interval speaks for parameter identifiability. Such behaviour is additionally confirmed by a moderate value of the mosquito-to-human ratio \( m \approx 3.385 \).

The histogram for the biting rate \( b_0 \) (Fig.~\ref{central_zone}$(b)$) shows a somewhat right-skewed distribution, having a clear peak at \( b_0 \approx 0.13 \) with a confidence interval leaning towards the higher side. It means that although the overall level of disease transmission is moderate, some instances exhibit a higher frequency of human-mosquito interactions. From an explanatory perspective, since the risk of infection is proportional to \( \lambda_H \sim b(t)\,m\, M_I / N_H \), a combination of low \( m \) and \( b_0 \) results in variations in transmission magnitude. Nonetheless, due to the sharpness of the histogram, instances where the contacts are high are rare occurrences, hence limiting significant outbreak magnification in this region.

The fraction of cases that show symptoms, \( p \) (Figure~\ref{central_zone}$(c)$), is surprisingly low, with a mean close to \( p \approx 0.028 \). Consequently, we observe that a very small proportion of cases are symptomatic, and there are quite many asymptomatic cases. As far as modelling is concerned, a small \( p \) allows for separating the measured incidence from the true level of infections, meaning that there could be more infections in reality. However, the narrow dispersion of \( p \), together with moderately high \( m \), makes the fitting procedure relatively robust.

Considering all this, it is safe to say that the Central zone features a regime of transmission where mosquitoes have a moderate density, contact rate, proportion with symptoms, and parameter variability. Such conditions ensure stability of dynamics, with no outbreak amplifications, which corresponds to narrow and steady forecast distribution patterns shown in Section 3.4.

\subsubsection*{West zone}
The fitted transmission dynamics of the West zone (Figure~\ref{west_zone}$(a)$) show that there exists a much higher regime of transmission when compared to that in the Central zone. It can be seen that the bootstrap confidence envelope is wider, and the median trend line shows a marked increase towards 2022 and then declines steadily up until 2024. Here, the inter-annual variation in the dataset (black dots) is greater and well captured by the model. The higher variability in the confidence envelope points to the higher sensitivity of this zone to changes in parameters, implying greater susceptibility to changes in the rate of vector density and contact. This is also reflected in the estimated mosquito-to-human ratio \(m \approx 3.397\).

In addition, it can be observed that the distribution of the baseline biting rate \( b_0 \) (Figure~\ref{west_zone}$(b)$) is more focused around \( b_0 \approx 0.10 \), showing a much narrower distribution when compared to that from the Central zone. It suggests that there is consistent maintenance of a moderately intense level of transmission in the West zone, as opposed to a scenario where intense transmission occurs due to sporadic occasions with high contact rate. In terms of mechanistic behavior, given that the force of infection is defined as \( \lambda_H \sim b(t)\,m\,M_I / N_H \), the increased value of \( m \) counterbalances the low value of \( b_0 \) leading to consistent transmission.

On the contrary, the symptomatic fraction \( p \) Fig.~\ref{west_zone}$(c)$ is noticeably higher than in the Central zone, with an average value of \( p \approx 0.054 \) and a more skewed distribution to the right. This implies that a higher number of cases will be symptomatic. Together with a relatively high number of mosquitoes, the stable biting rate, and the high level of symptomatic cases, this results in a transmission scenario characterized by notable and more easily observed outbreaks. This agrees with the predictions from the overall forecast distributions provided in the main text Section~3.4, which show a moderate future incidence increase for the West zone. In general terms, the West zone stands out as a system that has a stronger epidemiological footprint, with transmission being both persistent and effectively observed in the recorded number of cases.

\begin{figure}[ht]
	\centering
	\includegraphics[width = 1.0 \linewidth]{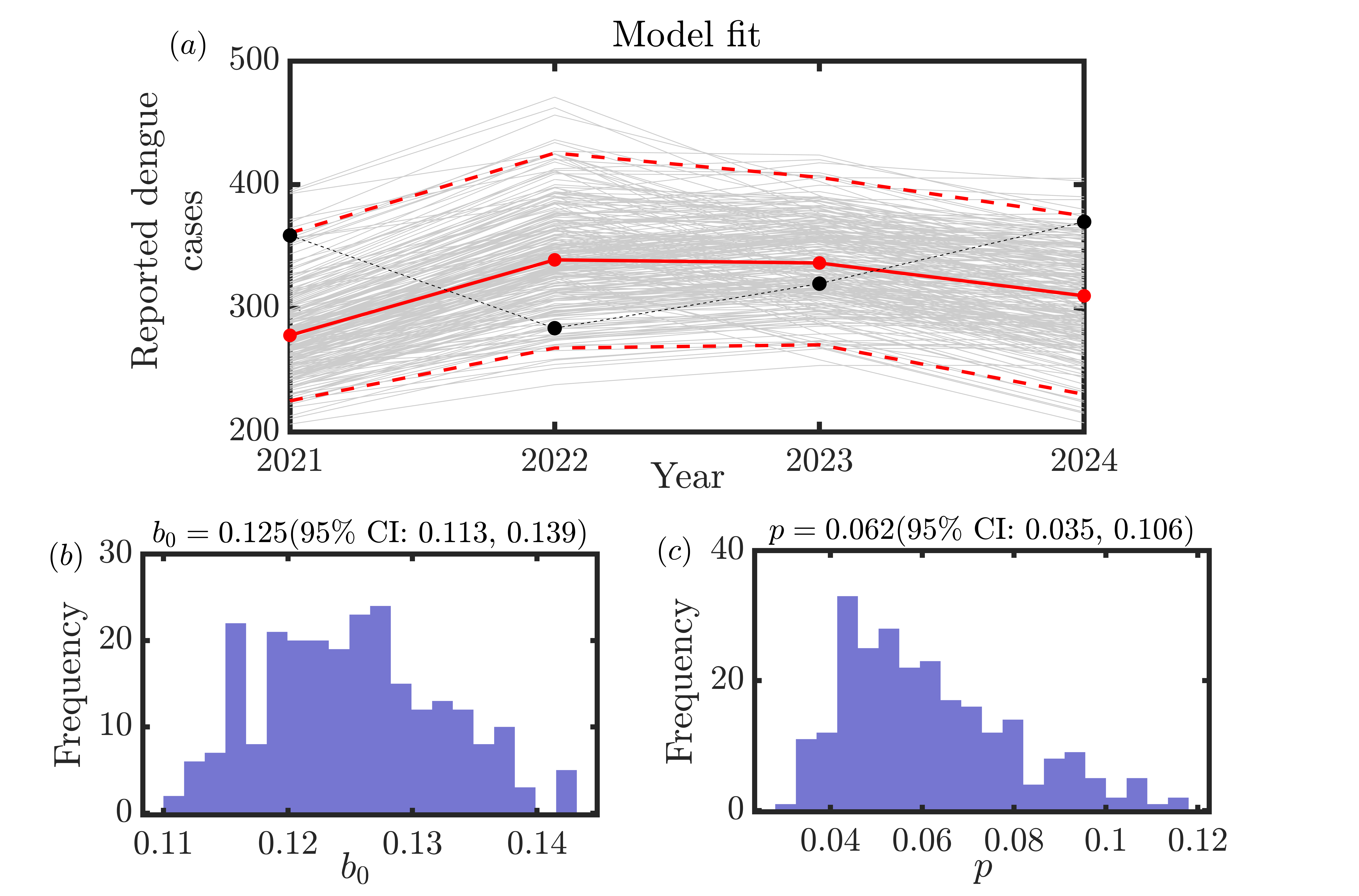} 
	\caption{\textbf{Northern region.} $(a)$ Reconstruction of a model of dengue cases with a regulated growth and subsequent stabilization for the entire period considered, with the bootstrap paths not exceeding the boundaries of a small confidence band. $(b)$ Distribution of the biting rate \( b_0 \) with values predominantly close to larger ones, representing a low variation of human-mosquito interaction rates. $(c)$ Distribution of the symptomatic fraction \( p \) with some variance and right skewness, revealing the heterogeneity of the symptomatic rate while maintaining a constant level of transmission activity.}
	\label{north_zone}
\end{figure}

\subsubsection*{North zone}
The North zone demonstrates a reasonably well-defined temporal development (Fig.~\ref{north_zone}$(a)$), where the number of cases rises in 2022 and then becomes gradually stabilized and even decreases slightly towards 2024. The set of bootstrap estimates stays reasonably well-bounded, suggesting that the reconstructed dynamics are reasonably robust against parameter variations. In contrast to other zones with higher variation, the North zone shows reasonable behavior in spite of having the same number of cases. This result agrees with the relatively small value of \( m \approx 2.021 \).

The distribution for the biting rate \( b_0 \) (Fig.~\ref{north_zone}$(b)$) appears centered on an appreciably large value but has a small and almost symmetric variance. Thus, human-mosquito interactions tend to occur at high levels constantly. Biologically, while a higher \( b_0 \) increases the infection pressure, the small value of \( m \) ensures that there is only regulation of outbreaks but not amplification. The small variance implies that this high level of human-mosquito interactions remains stable.

As can be seen in Fig.~\ref{north_zone}$(c)$, the symptomatic fraction \( p \) is relatively large and follows a skewed distribution to the right, which means that the share of infected individuals is high. The variability of the value of \( p \) when compared to the stability of the value of \( b_0 \) indicates that the variability is in relation to the manifestation of the disease and not in its transmission.

\begin{figure}[ht]
	\centering
	\includegraphics[width = 1.0 \linewidth]{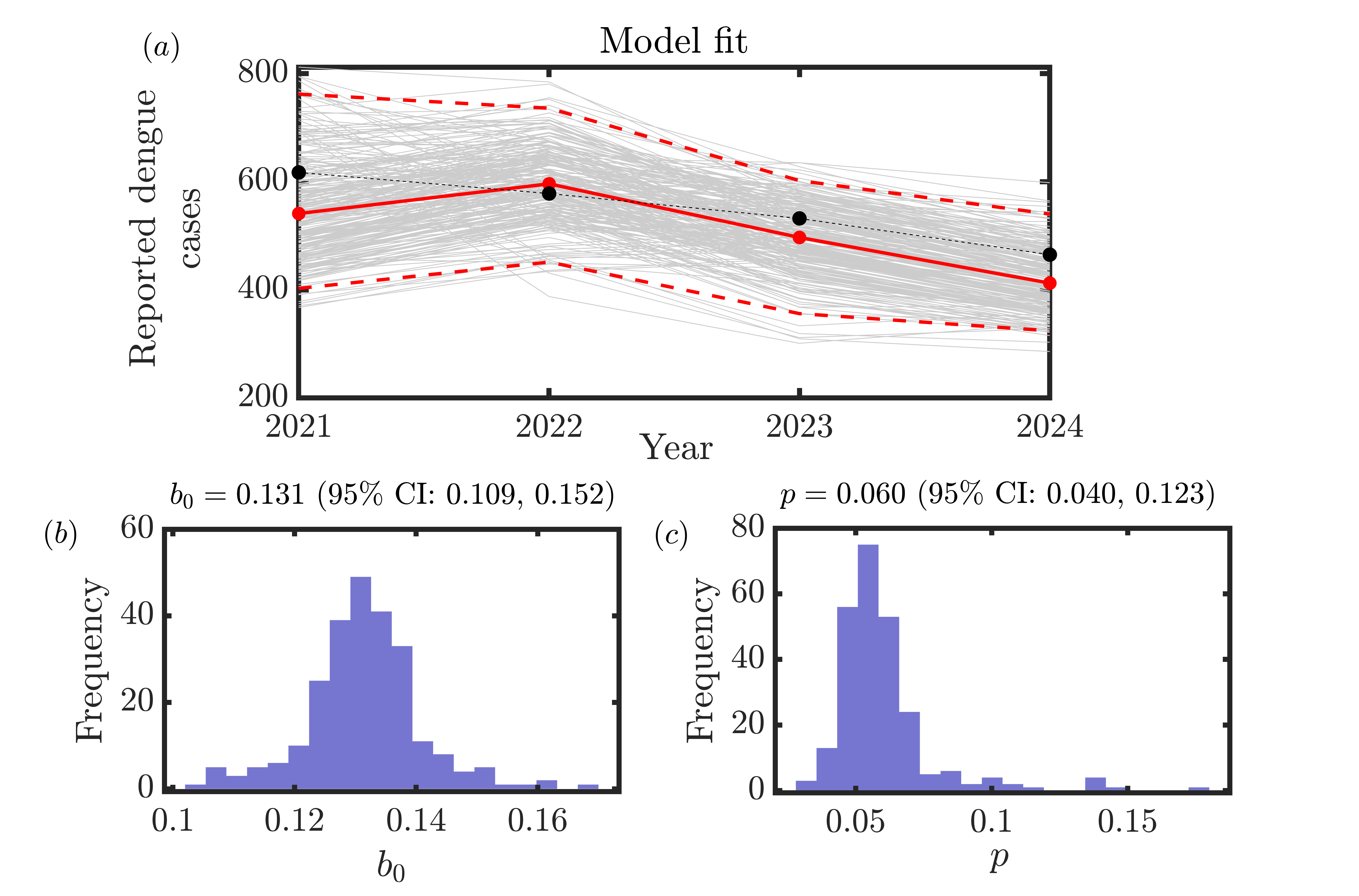} 
	\caption{\textbf{East zone.} $(a)$ Goodness of fit corresponding to a high incidence regime with a clear peak and subsequent decline, with a broad spread for the bootstrap distribution indicative of high variability. $(b)$ Distribution of the biting rate per unit time $b_0$, with a more spread-out distribution and high means, signifying variability in contact rate. $(c)$ Distribution of the proportion symptomatic $p$, which is right-skewed and has an extended tail.}
	\label{east_zone}
\end{figure}
\subsubsection*{East zone}

Unlike the more controlled zones, Zone East $(a)$ displays a high-intensity transmission profile characterized by a sharp rise in 2022, which is followed by a steady decrease up until 2024. The bootstrap ensemble is highly variable especially during the post-peak period due to the increased sensitivity of the model with diminishing transmission. The general intensity of the cases is much higher compared to other zones, showing that this zone is a significant source of the city-wide transmission dynamics. This happens despite the relatively low value of mosquito per person $m \approx 2.211$.

The baseline biting rate \( b_0 \) (Fig.~\ref{east_zone}$(b)$), on the other hand, has both high values and is widely distributed, thereby suggesting variation in the intensity of mosquito-human contact. This variation makes it possible for there to be a lot of variance in the force of infection and thus both an increase and a decrease in the epidemic size.

The fraction of symptomatic cases \( p \) (Fig.~\ref{east_zone}$(c)$) is relatively large and positively skewed. This long tail on the right side suggests a periodic increase in the number of symptomatic infections, potentially due to shifts in serotype circulation and immune responses. The combination of variations in \( b_0 \) and \( p \) along with a moderate value of \( m \) results in an intense and dynamic transmission mode, leading to extensive outbreaks.

\begin{figure}[h]
	\centering
	\includegraphics[width = 1.0 \linewidth]{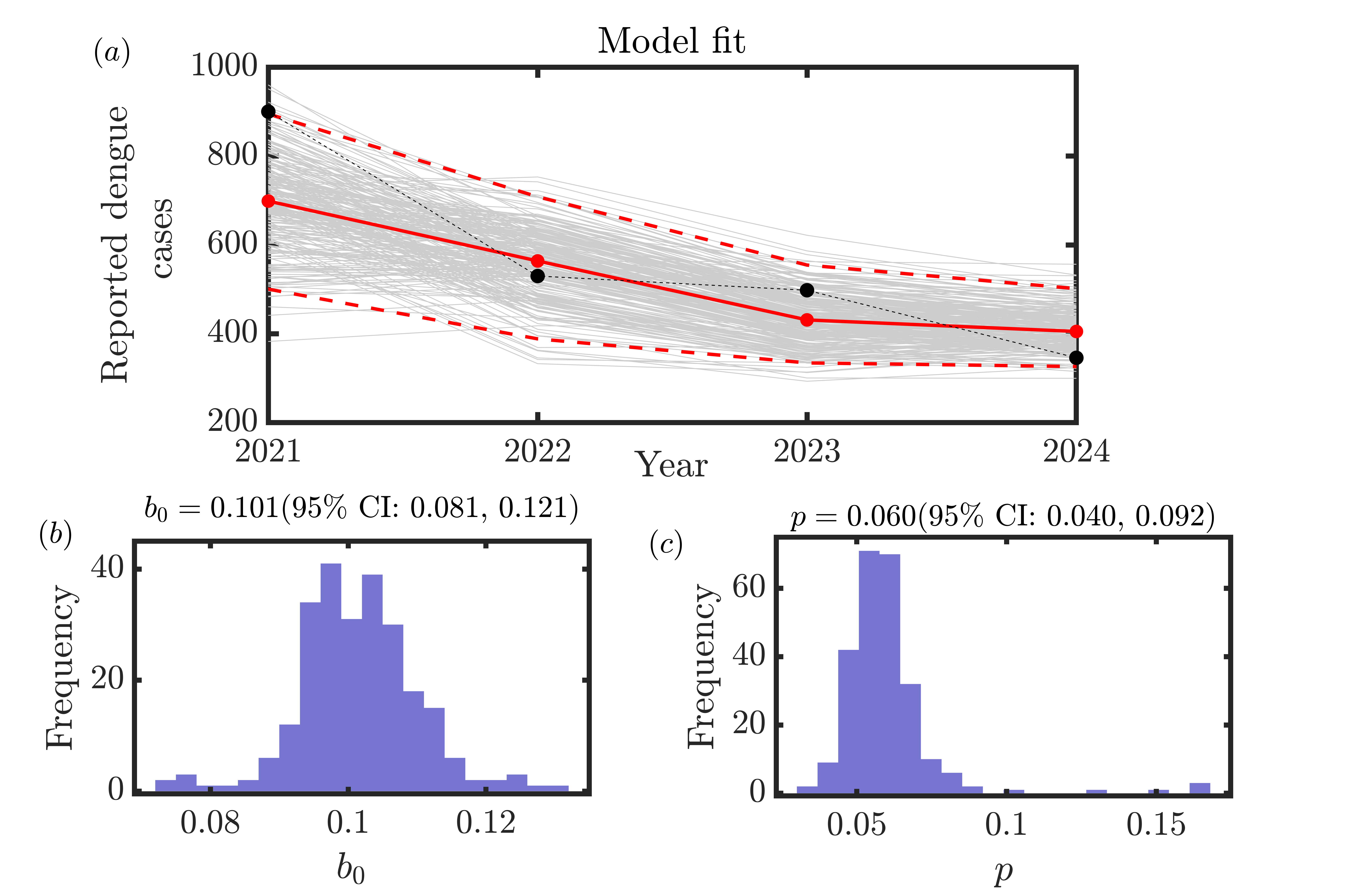} 
	\caption{\textbf{South zone}. $(a)$ Model fit depicting a high initial disease burden, then steady decrease in dengue cases reported, where the bootstrap sample replicates the general trend even when variability is fairly large. $(b)$ Baseline biting rate \( b_0 \) distribution, which shows moderate spread but symmetric distribution and thus indicates that contact rates are stable. $(c)$ Symptomatic fraction \( p \) distribution, which is moderately distributed, and has a right skew with low variability.}
	\label{south_zone}
\end{figure}

\subsubsection*{South zone}
High initial load with a continuous reduction pattern is typical of the South zone (see Fig.~\ref{south_zone}$(a)$). The median profile shows the continuous decreasing trend, while the bootstrap profiles retain a wide but well-defined shape. This shows that the process is stable under different parameter instances. The continuous decrease over time means that the process will be in a high load state and will move towards stabilization. This type of behavior is significantly determined by the high mosquito to human ratio \( m = 4.620 \).

The baseline biting rate \( b_0 \) (Fig.~\ref{south_zone}$(b)$) exhibits a moderate value with a narrow spread, suggesting stable human-mosquito interaction intensities among realizations. In this scenario, variation in transmission does not arise from variations in the intensity of interactions but rather from the overwhelming effect of vector density. This can be attributed to the large value of \( m \), which increases the effective force of infection despite the moderate \( b_0 \).

The symptomatic parameter \( p \) (Fig.~\ref{south_zone}$(c)$) is not small but is slightly right-skewed, implying that there is a reasonable fraction of cases that can be observed symptomatically. The skewness to the right shows that, at some instances, the expression of the symptomatic parameter has increased, perhaps due to differences in immune responses or reporting. The dynamics in the South zone are mainly determined by the vector population, such that a large \( m \) leads to strong infection rates.

Collectively, the South zone is characterized by a transmission scenario in which vector abundance is the dominant force driving epidemics, with contact intensity and symptomatic fractions remaining moderate. Here, the high value of \( m \) is the main determinant of epidemic severity, allowing for large burdens of cases even when \( b_0 \) remains moderate. This matches the forecasting behavior highlighted in Section~3.4, whereby the South zone exhibits an ongoing trend of steadily declining case counts. As such, from a public health standpoint, this suggests that vector control efforts, particularly those aimed at managing larvae and their environment, will have a far greater impact than approaches that attempt to reduce only human-vector interactions.

\begin{figure}[h]
	\centering
	\includegraphics[width = 1.0 \linewidth]{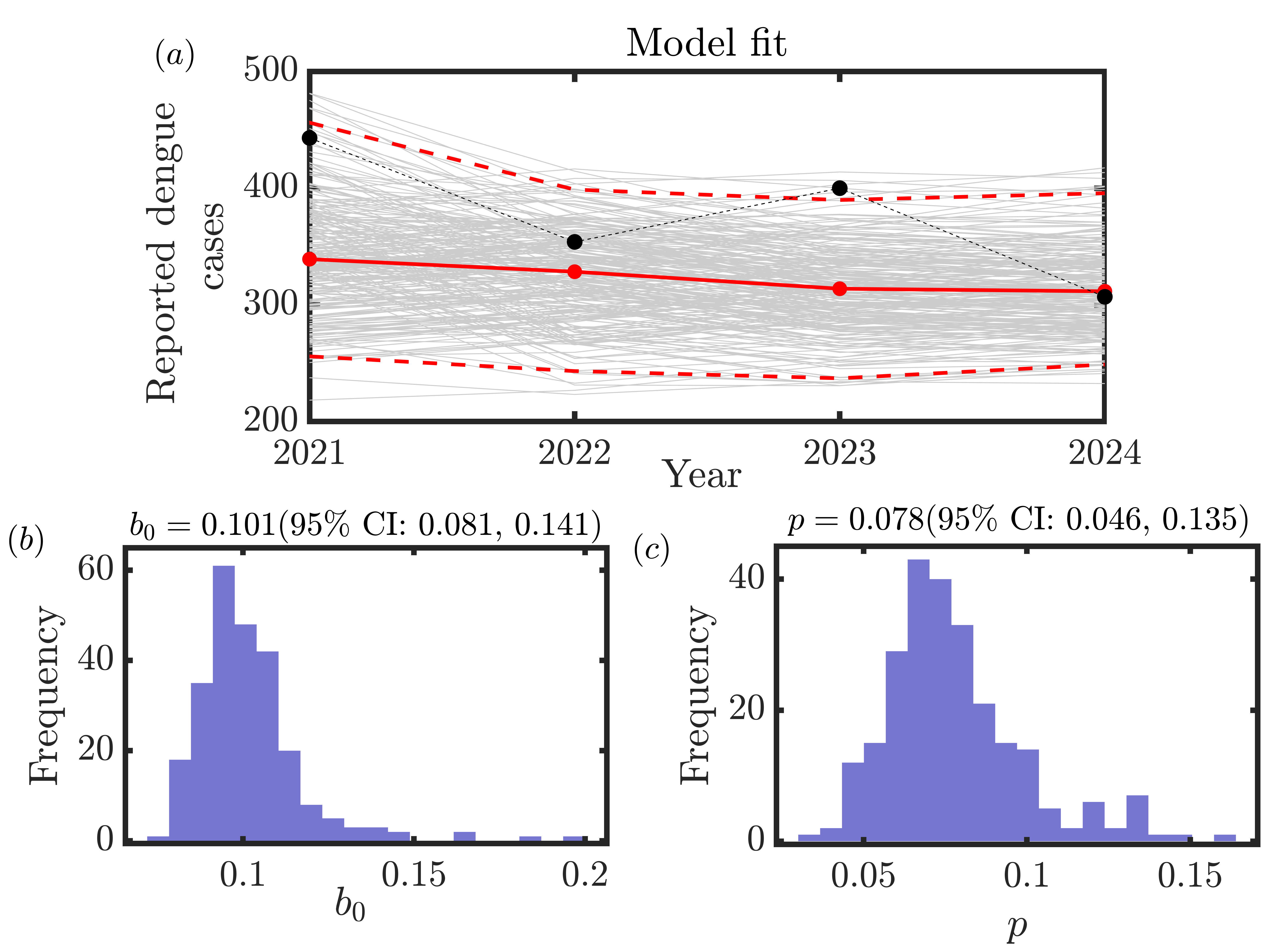} 
	\caption{\textbf{North-west zone.} $(a)$ Temporal dynamics of model fitness that show non-monotonicity with periods of decay followed by revival, together with a broad variability of bootstrap paths which reflects strong dependency on parameter changes. $(b)$ Probability distribution of \( b_0 \), a baseline biting rate value. The value is moderate on average, but a relatively broad distribution indicates significant variability of transmission intensity. $(c)$ Probability distribution of \( p \), a symptomatic fraction; high mean with a broad variability.}
	\label{north_west}
\end{figure}

\subsubsection*{North-west zone}
A high degree of temporal variability is evident in the North-West region (see Fig.~\ref{north_west}$(a)$), where there is an initial drop in infection numbers, followed by an increase in 2023 and another drop in 2024. Such non-monotonic dynamics are well-represented by the fitted curve, but the large dispersion among the bootstrap sample members highlights a considerable dependence on parameter values. In comparison with the other zones, oscillations appear more frequent, implying that there is dynamical instability within the transmission process. This correlates well with the largest mosquito-to-human ratio \( m \approx 4.815 \).

The biting rate \( b_0 \) (Fig.~\ref{north_west}$(b)$) is moderately high but shows a fairly wide range of variation, reflecting differences in contact frequency in actual cases. A modest level of variation in \( b_0 \) can have a substantial effect on the infection rate due to the high value of \( m \). This combination of contact variation and vector abundance is responsible for the observed oscillation in the incidence rate.

The symptomatic fraction \( p \) (see Fig.~\ref{north_west}$(c)$) is rather large and widely spread. As a result, the presence of very large values of \( m \), combined with varied values of \( b_0 \) and rather large values of \( p \), results in an explosive transmission environment.

The combination of the very high mosquito population and the relatively high symptomatic rate leads to transmission that is very intensive and highly noticeable but variable. This accounts for the variable behavior observed in the fitted curves and is reflected in the increased and broader forecast distribution presented in Section~3.4. For public health practice, the North-west zone becomes an important hot spot, where the presence of many vectors and efficient case detection contribute to continued transmission. In this zone, there is a need for vector control and surveillance, considering the high value of \( m \).

\begin{figure}[h]
	\centering
	\includegraphics[width = 1.0 \linewidth]{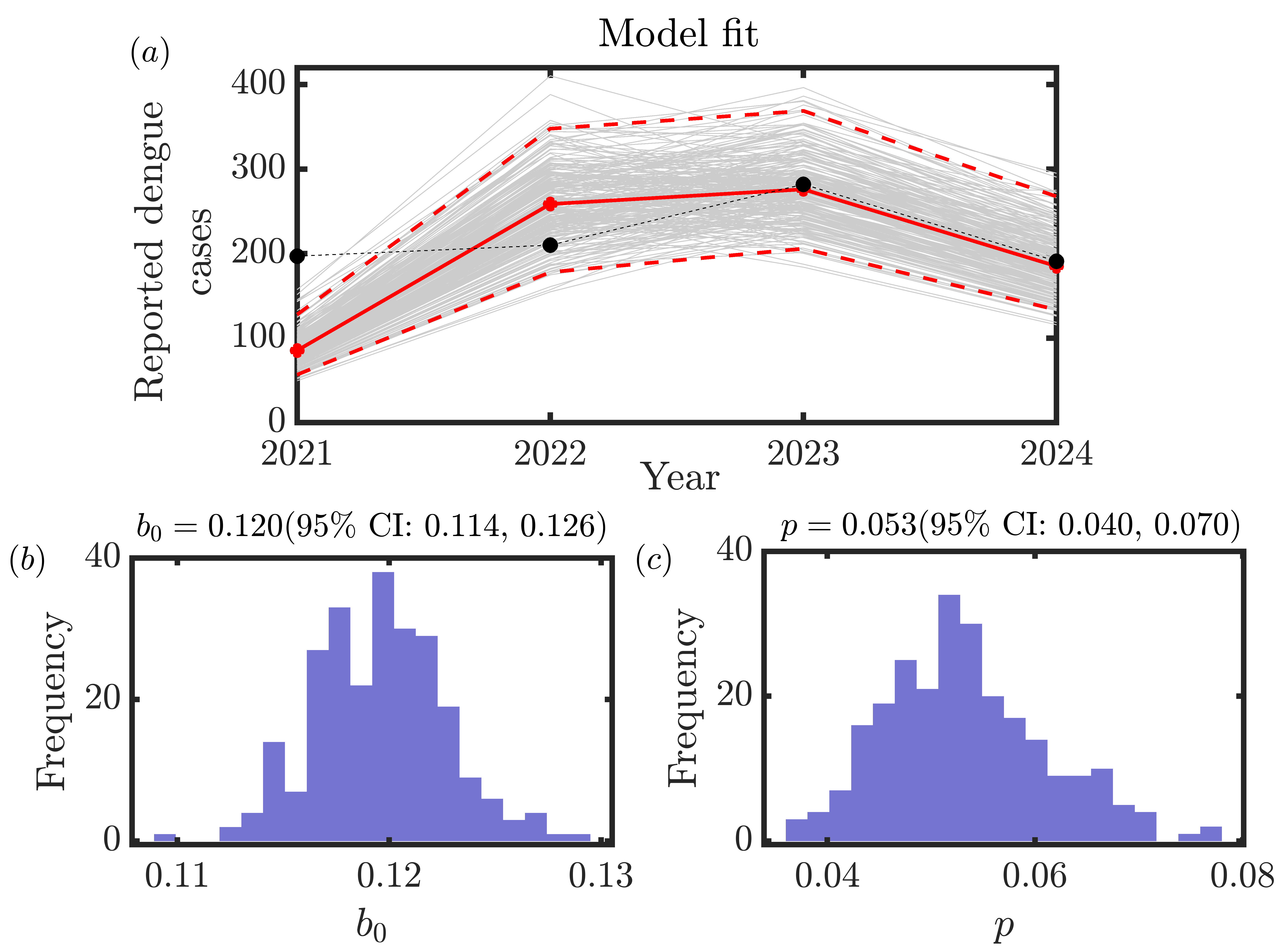} 
	\caption{\textbf{South-west zone.} $(a)$ Model goodness of fit, where the cases increase gradually and stabilize, after which they start falling, while the bootstrap distributions show moderate variability around the median curve. $(b)$ Biting rate distribution \( b_0 \), centered on moderate values, indicative of stable human-mosquito contacts. $(c)$ Symptomatic proportion \( p \) distribution, with mild right skewness.}
	\label{south_west}
\end{figure}

\subsubsection*{South-west zone}
The trend for the South-west zone (Fig.~\ref{south_west}$(a)$) is characterized by an increase in cases followed by a slow decrease. The median profile reflects this trend from an increase to stabilization, and the bootstrap ensemble shows an intermediate range, suggesting that the trend is neither stable nor unstable but somewhere in between. Contrary to zones with high amplification and regulation, the trend in this case remains moderate throughout the whole study period. This trend can be attributed to an intermediate mosquito-to-human ratio \( m = 3.0 \).

The density of \( b_0 \) (Fig.~\ref{south_west}$(b)$) is highly concentrated around a moderate value, signifying constant contact rate between humans and mosquitoes in all the realizations. The consistency restricts the variation in transmission intensity, thereby causing the smooth transition seen in the fitted curves. The balance in the force of infection arises due to a moderate value of \( b_0 \) along with the intermediate value of \( m \).

The symptomatic ratio \( p \) (Fig.~\ref{south_west}$(c)$) is relatively high, implying that there may be a fair number of infections which are symptomatic. Since there is no significant variability in the parameters \( b_0 \) and \( p \), it is safe to conclude that the stability of the system persists. As a result, the South-west region can be considered an intermediate transmission scenario, where neither vectors nor contacts play a major role.

\medskip

In general, through the zone-wise analysis of the data, we learn that the heterogeneity in the future dengue outbreak in Ahmedabad city results from the emergence of distinct regimes of the underlying dynamics, which can be explained by the interaction of $m$, $b_0$, and $p$. Regions with high-$m$ values (South and North-West zones) produce highly intensified and unpredictable dengue outbreaks; moderate-$m$ regions (Central, West, South-West zones) are associated with regulated dynamics characterized by different degrees of sensitivity; finally, low-$m$ regions (North and East zones) rely strongly on the level of contacts.

\section*{Zone-wise impact of intervention strategies}
As an extension to the zonal model fitting and forecasting dynamics, we assess the efficiency of vector control strategies implemented in all municipal zones. With respect to the same set-up of intervention as for AMC, we analyze the impact that seasonal suppression of mosquito population can have on the number of dengue infections. The analysis is performed with respect to two types of interventions, which differ in terms of time span of their application, enabling us to compare short-time versus sustained suppression effect. Of crucial importance here is the zonal response in terms of transmission intensity \(m\).

\begin{table}[h!]
	\centering
	\caption{Zone-wise percentage reduction in reported dengue cases under 4-month fogging intervention (June--September).}
	\begin{tabular}{lccc}
		\hline
		\textbf{Zone} & \textbf{2026 (\%)} & \textbf{2027 (\%)} & \textbf{2028 (\%)} \\
		\hline
		Central     & 20.42 & 55.66 & 62.48 \\
		West        & 17.38 & 63.25 & 79.80 \\
		North       & 22.98 & 66.48 & 83.27 \\
		East        & 16.56 & 62.69 & 78.42 \\
		South       & 19.85 & 60.24 & 71.10 \\
		North-West  & 20.50 & 58.92 & 68.81 \\
		South-West  & 7.97  & 58.04 & 74.62 \\
		\hline
	\end{tabular}
	\label{tab:fogging}
\end{table}
\subsection*{Fogging (4 months: June--September)}

Fogging short-term effects can be presented in the table below (Table~\ref{tab:fogging}). As we can see, there is a pronounced temporal gradient in each of the zones, where the decrease in values is moderate (about 16-23\% decrease in 2026) and continues until 2027 and 2028, when it significantly increases. Thus, for example, in the North zone, the increase is from 22.98\% (2026) to 83.27\% (2028). In the West zone, it is from 17.38\% to 79.80\%.

Zone-wise comparison analysis has shown that heterogeneous effects are dependent on the strength of transmission within the respective zones. Zones that have relatively weak interaction strengths, such as East with \(m = 2.211\), take a longer time to reduce infections to 16.56\% by 2026. However, other zones, including North with \(m = 2.021\) and West with \(m = 3.397\), show a greater response effect in later years. Surprisingly, zones with higher values of \(m\), such as South with \(m = 4.620\) and North-West with \(m = 4.815\), do not have the highest percentage reduction immediately, but they continue to improve steadily, having reached 71.10\% and 68.81\%, respectively, by 2028.

However, the south-western region still achieves the least reduction initially (7.97\% by 2026) because of its late outbreak pattern shown in the prediction. Nonetheless, the rate is expected to spike to 74.62\% by 2028, implying that repeated season interventions will eventually compensate for the lower initial response rate. Thus, Table~\ref{tab:fogging} shows that fogging serves as a slow-down process, where success is heavily reliant on time and the transmission process.

\begin{table}[h!]
	\centering
	\caption{Zone-wise percentage reduction in reported dengue cases under 6-month spraying intervention (June--November).}
	\begin{tabular}{lccc}
		\hline
		\textbf{Zone} & \textbf{2026 (\%)} & \textbf{2027 (\%)} & \textbf{2028 (\%)} \\
		\hline
		Central     & 26.63 & 74.75 & 85.54 \\
		West        & 24.05 & 79.85 & 93.69 \\
		North       & 29.16 & 81.57 & 94.73 \\
		East        & 23.24 & 79.48 & 93.18 \\
		South       & 26.16 & 77.98 & 90.63 \\
		North-West  & 26.76 & 77.11 & 89.15 \\
		South-West  & 14.98 & 77.02 & 91.99 \\
		\hline
	\end{tabular}
	\label{tab:spraying}
\end{table}

\subsection*{Spraying (6 months: June--November)}
Effectiveness of prolonged spraying is demonstrated in Table~\ref{tab:spraying}, and the difference between fogging can be easily observed. In all the zones, declines of infection levels in 2026 have been more pronounced than during fogging and exceed 23\%. As an illustration, for the North zone, the decline equals 29.16\%, which is more than the 22.98\% drop observed during fogging. The Central zone demonstrates a similar trend, rising from 20.42\% to 26.63\%.

However, the impact is even more profound in the long term. The decline in prevalence is more than 75\% across all zones by 2027 and 2028, while in certain critical zones, it reaches or even exceeds 90\%. Particularly, the North zone (94.73\%), West zone (93.69\%), and East zone (93.18\%) experience close-to-complete elimination by 2028, indicating the susceptibility of heavily burdened systems to extended vector elimination efforts.

The importance of the transmission parameter \(m\) is better understood when considering comparisons between different zones. Zones with high \(m\) values, like South and North-West, have achieved a huge improvement compared to fogging, with reductions of 90.63\% and 89.15\%, respectively, up to 2028. This proves that prolonged intervention period works well to neutralize the very strong self-amplification factor within the zones. The South-West region, on the other hand, which recorded low effectiveness when using fogging, has shown remarkable improvement with a reduction rate of 91.99\% up to 2028.

On the whole, Table~\ref{tab:spraying} demonstrates that the impact of spraying goes beyond being simply a modification of fogging and provides an improved method for epidemic control. By prolonging the period of the applied measures, it reduces the transmission not only during its peak but also stops the emergence of a resurgence phase, resulting in a rapid drop on the entire territory.

When viewed collectively, the results from the zonal modelling, forecasting, and intervention studies give an integrated view of the dengue transmission dynamics within a heterogeneous urban environment. Through the estimated model parameters, it becomes apparent that the intrinsic transmission force, measured using \(m\), controls not only the basic reproductive value of the disease but also the sensitivity of each zone to different types of interventions. Although short-term interventions like fogging lead to incremental reductions, they ultimately have an inherent limitation of efficacy in areas with persistent levels of transmission force. On the other hand, long-term interventions like spraying prove effective in reducing peak transmission and its re-emergence, resulting in almost complete suppression.

The most important aspect of the findings is that it clearly illustrates that effectiveness of interventions depends on the local dynamical structure and hence an overarching strategy may not be effective in urban environments in managing epidemics. In contrast, with the use of data-driven models and region-specific forecasts, it would be possible to plan interventions such that the timing and period of the intervention are optimally decided. Hence, this research provides us with a quantitatively sound framework based on epidemiological data, dynamic modelling, and interventions in tackling dengue epidemics.

\nolinenumbers



\section*{Acknowledgments}
SR and IG are supported by the Gates Foundation (INV-044445). IG is also supported by ARG-MATRICS (ANRF/ARGM/2025/000676/MTR).

\nolinenumbers

%
%
%
\bibliography{bibliography}

\end{document}